\documentclass[%
superscriptaddress,
 amsmath,amssymb,
 aps,
pre,
10pt,
longbibliography, citesort
]{revtex4-1}
\usepackage{color}
\usepackage{graphicx}
\usepackage{dcolumn}
\usepackage{bm}
\usepackage{chemarrow}
\usepackage{subfigure}
\usepackage{epstopdf}
\usepackage{CJK}

\begin{document}
\begin{CJK*}{UTF8}{gbsn}

\title{Fluctuations and correlations of reactive scalars near chemical equilibrium \\in incompressible turbulence}

\author{Wenwei Wu (吴文伟)}
 \email{wenwei.wu@etu.univ-littoral.fr}
 \affiliation{Univ. Lille, ULR 7512 - Unit\'e de M\'ecanique de Lille - Joseph Boussinesq (UML), F-59000 Lille, France}%
 \affiliation{CNRS, Univ. Lille, ULCO, Laboratory of Oceanology and Geosciences, UMR LOG 8187, Wimereux, France}%
 \affiliation{UM-SJTU Joint Institute, Shanghai Jiao Tong University 200240, Shanghai, China}%


\author{Enrico Calzavarini}
\affiliation{Univ. Lille, ULR 7512 - Unit\'e de M\'ecanique de Lille - Joseph Boussinesq (UML), F-59000 Lille, France}%
\author{Fran\c{c}ois G. Schmitt}
\affiliation{CNRS, Univ. Lille, ULCO, Laboratory of Oceanology and Geosciences, UMR LOG 8187, Wimereux, France}%

\author{Lipo Wang (王利坡)}%
\affiliation{UM-SJTU Joint Institute, Shanghai Jiao Tong University 200240, Shanghai, China}%

\date{\today}

\begin{abstract}
The statistical properties of species undergoing chemical reactions in a turbulent environment are studied. We focus on the case of reversible multi-component reactions of second and higher orders, in a condition close to chemical equilibrium sustained by random large-scale reactant sources, while the turbulent flow is highly developed. In such a state a competition exists between the chemical reaction that tends to dump reactant concentration fluctuations and enhance their correlation intensity and the turbulent mixing that on the contrary increases fluctuations and remove relative correlations.
We show that an unique control parameter, the Damkh\"{o}ler number ($Da_\theta$) that can be constructed from the scalar Taylor micro-scale, the reactant diffusivity and the reaction rate characterizes the functional dependence of fluctuations and correlations in a variety of conditions, i.e., at changing the reaction order, the  Reynolds and the Schmidt numbers.
The larger is such a Damkh\"{o}ler number the more depleted are the scalar fluctuations as compared to the fluctuations of a passive scalar field in the same conditions, and vice-versa the more intense are the correlations. A saturation in this behaviour is observed beyond $Da_\theta \simeq \mathcal{O}(10)$. We provide an analytical prediction for this phenomenon which is in excellent agreement with direct numerical simulation results.
\end{abstract}


\maketitle
\end{CJK*}

\section{Introduction}\label{Introduction}
Turbulence is an ubiquitous complex phenomenon, found in many applied situations such as
the automotive industries, chemical engineering, the environment including astrophysics, meteorology and oceanography.
In fully developed turbulence, universal scaling relations are classically discussed in the framework of
Kolmogorov-Obukhov phenomenology for the velocity field, as well as Obukhov-Corrsin approach for
passive scalars advected by the turbulent velocity. This now forms the classical KOC (Kolmogorov-Obukhov-Corrsin)
theoretical framework for fluid turbulence with advected passive scalars \citep{warhaft2000passive}.

Chemical reactions have been considered quite early in such framework, for some specific cases.
For example, \citet{Corrsin1961} studied the mixing of a scalar contaminant undergoing a first-order chemical reaction in isotropic turbulence. Theoretically he deduced the power spectrum of the reactive scalar in different wave number ranges. Later, \citet{pao1964statistical} investigated the dilute turbulent concentration fields of a multicomponent mixture with first-order reaction and proposed an unified spectral transfer concept for deducing the scalar spectrum at large wave numbers. For higher-order reactions, nonlinearity from the chemical source have also been introduced: \citet{obrien1966,obrien1971} worked on decaying second-order, isothermal reaction in turbulence. The asymptotic decay rate of the scalar energy was found as $t^{-11/2}$ for moderate reactions, and $t^{-3/2}$ for rapid reactions. The covariance between reactants and the development
of models for the covariance terms were important topics of turbulent mixing analyses as well~\citep{Lamb1978,Heeb1990Turbulent}. For the two non-premixed reactants case, it was found
\citep{toor1969turbulent} that the covariance is almost invariant for very slow and very rapid second-order reaction.

Most recent work \textcolor{black}{on} such topic have been done for fast reactions and combustion as application field.
Here we mostly focus on higher order reactions, with possible applications in the field of chemical and biological oceanography,
where the typical times of biogeochemical reactions may be large. In such context, some results
on the phytoplankton statistics, considered via the proxy of fluorescence measurements, have found
some scaling relations with $-1.2$ spectral slopes, interpreted as signature of biological activity \citep{Seuront2013Multifractal,Lovejoy2001,Derot2015}. This generated our motivation here to study
the statistical properties of various high-order reactions in a turbulent flow.

In the present work we focus on the fundamental properties of reactive scalar mixing in incompressible turbulence. The flow statistics, from global to scale-dependent features are studied in details and theoretically modelled. The article is organized as follows: In section \ref{problem}, we introduce the model system, its governing equations together with the set of dimensionless control parameters. Section \ref{DNS} briefly details the numerical methods adopted in this study. The results and their analyses are described in section \ref{results}. Finally, section~\ref{conclusion} summarizes the main findings, discuss their implications and future perspectives.

\section{Problem definition} \label{problem}
In this study we consider reactions of the form:
\begin{equation}
\textrm{R}_1+n\textrm{R}_2\autorightleftharpoons{$_{\gamma_1}$}{$^{\gamma_2}$} \textrm{P},
\end{equation}
where $\textrm{R}_1$, $\textrm{R}_2$ and $\textrm{P}$ denote three generic reactive scalars 
\textcolor{black}{where $\textrm{R}_1$, $\textrm{R}_2$ and $\textrm{P}$ denote three generic reactive scalars and $n$ is an integer coefficient. The process is reversible with independent non-zero forward/backward reaction rates $\gamma_1$ and $\gamma_2$. The order of the chemical reaction, which is defined as the sum of the powers of the reactants' concentration  in the rate equation is $n+1$ for the forward reaction, because the rate equation reads $\gamma_1 R_1R_2^n$,  while is of the first order for the backward reaction with reaction rate $\gamma_2 P$. Moreover, the reactants are assumed to be subject to molecular diffusion and to fluid advection.}

The evolution equations for the concentration fields $R_1(\mathbf{x},t)$, $R_2(\mathbf{x},t)$ and $P(\mathbf{x},t)$ read:
\begin{eqnarray} 
  \label{eq_flow} \partial_t \mathbf{u}+(\mathbf{u}\cdot\nabla)\mathbf{u}&=&\nu\triangle\mathbf{u}-\nabla p/\rho+\mathbf{f},\\
 \label{eq_mass} \textcolor{black}{\nabla\cdot\mathbf{u}}&\textcolor{black}{=}&\textcolor{black}{0,}\\
  \label{eq_scalar_1} \partial_t R_1+(\mathbf{u}\cdot\nabla)R_1&=&D\triangle R_1-\gamma_1 R_1R_2^n+\gamma_2 P+\dot{q}_{R_1},\\
  \label{eq_scalar_2} \partial_t R_2+(\mathbf{u}\cdot\nabla)R_2&=&D\triangle R_2-n(\gamma_1 R_1R_2^n-\gamma_2 P)+\dot{q}_{R_2},\\
  \label{eq_scalar_3} \partial_t P+(\mathbf{u}\cdot\nabla)P&=&D\triangle P+\gamma_1 R_1R_2^n-\gamma_2 P+\dot{q}_P.
\end{eqnarray}
Here $\mathbf{u}(\mathbf{x},t)$ is \textcolor{black}{an incompressible} three-dimensional flow velocity \textcolor{black}{described by the Navier-Stokes equations (\ref{eq_flow})-(\ref{eq_mass})}, $p$ is the pressure, $\rho$ is the fluid density set as constant, $\nu$ is the kinematic viscosity and $D$ is the species diffusivity (assumed as being the same for all species).
To sustain the turbulent fluctuations, large-scale forcing terms $\mathbf{f}$ and $\dot{q}$ are introduced for the velocity and scalars, respectively. More details about the expression of these forcing terms will be provided in section \ref{DNS}.

For comparison, a non-reactive species $\textrm{T}$ undergoing both advection and diffusion is also considered in this study. Its local concentration evolves according to the following equation,
\begin{align}
  \label{eq_scalar_4} &\partial_t T+(\mathbf{u}\cdot\nabla)T=D\triangle T+\dot{q}_T.
\end{align}

The equations for the above model system can be made dimensionless by choosing reference scales appropriate for the present system.
Since the turbulent flow is unbounded, we take the Taylor microscale ($\lambda$) and the single component velocity fluctuation ($u'$) as the reference scales for space and velocity, respectively, which are defined as:
$\lambda=\sqrt{\frac{15\nu}{\varepsilon}}u',\qquad
u'=\frac{1}{3} \sum_{i} \sqrt{\langle u_i^2 \rangle},\qquad  \varepsilon =  \frac{\nu}{2} \langle\sum_i \sum_j (\partial_i u_j + \partial_j u_i)^2 \rangle. $
Here $\varepsilon$ is the mean dissipation rate and $\langle \ldots \rangle$ denotes the space and time (or equivalently ensemble) average.
The scalar quantities can be non-dimensionalized by means of their equilibrium values in no-flow conditions $R_{1,eq}, R_{2,eq}, P_{eq}$, while for the passive scalar the global mean $\langle T \rangle$ is used as the reference value.
Note that at the equilibrium, the algebraic relation $\gamma_1 R_{1,eq}R_{2,eq}^n=\gamma_2 P_{eq}$ holds.
Furthermore, in the present work for simplicity we assume that $R_{1,eq}=R_{2,eq}$.
This leads to the following dimensionless equations:
\begin{eqnarray}
  \label{eq_flow_non_dim_used} \partial_t \mathbf{u}+(\mathbf{u}\cdot\nabla)\mathbf{u}&=&Re_{\lambda}^{-1}\triangle\mathbf{u}-\nabla p+\mathbf{f},\\
  \label{eq_mass} \textcolor{black}{\nabla\cdot\mathbf{u}}&\textcolor{black}{=}&\textcolor{black}{0,}\\
  \label{eq_scalar_non_dim_1_used}\partial_t R_1+(\mathbf{u}\cdot\nabla)R_1&=&(Sc\ Re_{\lambda})^{-1}\triangle R_1-Da (R_1R_2^n-P)+\dot{q}_{R_1},\\
  \label{eq_scalar_non_dim_2_used}\partial_t R_2+(\mathbf{u}\cdot\nabla)R_2&=&(Sc\ Re_{\lambda})^{-1}\triangle R_2-nDa (R_1R_2^n-P)+\dot{q}_{R_2},\\
  \label{eq_scalar_non_dim_3_used}\partial_t P+(\mathbf{u}\cdot\nabla)P&=&(Sc\ Re_{\lambda})^{-1}\triangle P+Da (R_1R_2^n-P)+\dot{q}_P,\\
  \label{eq_scalar_non_dim_4_used}\partial_t T+(\mathbf{u}\cdot\nabla)T&=&(Sc\ Re_{\lambda})^{-1}\triangle T+\dot{q}_T,
\end{eqnarray}
where $Re_{\lambda}=\lambda\cdot u'/\nu$ is the Taylor based Reynolds number, the Schmidt number $Sc=\nu/D$ is the ratio of viscous diffusion to molecular diffusion, the Damkh\"{o}ler number $Da=\lambda \gamma_1 R_{2,eq}^n/u' = \lambda \gamma_2/u'$ represents the ratio of flow timescale to the chemical timescale of forward or backward reaction. Note that the particular choice $R_{1,eq}=R_{2,eq}$ is crucial in obtaining a single Damkh\"{o}ler number, instead of two distinct ones that would be present in general cases.

In conclusion the control parameters of the model system are $Re_{\lambda}$, $Sc$, $n$ and $Da$.

\section{Numerical methods}\label{DNS}
The model system presented in Sec.\ref{problem} is numerically simulated in a cubic tri-periodic domain.
The flow is sustained by a large-scale forcing capable to generate a statistically steady homogeneous and isotropic turbulent flow. The expression of the forcing field in Fourier space, $ \hat{\mathbf{f}}(\mathbf{k},t)$ reads,
\begin{equation}\label{f_dot}
   \hat{\mathbf{f}}(\mathbf{k},t)=\frac{1}{\tau_f}\sum_{1\leqslant|\mathbf{k}|\leqslant2\sqrt{2}}\hat{\mathbf{u}}(\mathbf{k,t}),
\end{equation}
with $\tau_f$ a time-scale being adjusted at each time step in order to provide a constant power input, i.e. $ \int_V \mathbf{f}\cdot\mathbf{u} dx^3= const. $. This type of forcing, called linear, has been adopted e.g. in \citep{Schumacher2007}. Note also that the zero mode $|\mathbf{k}|=0$ is not forced in order to prevent the development of a global mean flow, i.e., in our simulations $\langle \mathbf{u} \rangle = 0$.
Similarly, the external source term on scalars ($\dot{q}_{\theta}$ with $\theta=R_1,R_2,P$ or $T$ in Eq.~\eqref{eq_scalar_non_dim_1_used}-\eqref{eq_scalar_non_dim_4_used}) is also isotropic and acting at the largest scales; however it is constant in amplitude~\citep{alvelius1999random,Gotoh2011,Gotoh2015}. In Fourier space this reads,
\begin{equation}\label{q_dot}
   \hat{\dot{q}}_{\theta}(t)=\sum_{1\leqslant|\mathbf{k}|\leqslant2\sqrt{2}}\frac{Q}{|\mathbf{k}|}e^{i\phi_{\theta}(\mathbf{k},t)},
\end{equation}
where $Q$ is the constant prescribing the overall source amplitude, $|k|^{-1}$ is a normalization factor to guarantee that the forcing amplitude is larger at small wave numbers. In particular, the random phase function $\phi_{\theta}(\mathbf{k},t)$ is generated independently for each scalar field and delta-correlated both in time and in wave-vector~\citep{Gotoh2011,Gotoh2015}. As a result, $\dot{q}_{R_1}$, $\dot{q}_{R_2}$, $\dot{q}_P$ and $\dot{q}_T$ have amplitudes of the same order, but they are statistically independent from each other both in time and in space.
The set of dynamical equations ~\eqref{eq_flow_non_dim_used}-\eqref{eq_scalar_non_dim_4_used} are solved numerically by means of a pseudo-spectral code~\cite{gauding2017high,gauding2018one} using a smooth dealiasing technique for the treatment of non-linear terms in the equations~\citep{Hou2007}. Compared with the conventional $2/3$ rule approach, such smooth dealiasing is capable to reduce numerical high frequency instabilities. The time-marching scheme adopts a third order Runge-Kutta method.

We explore the parameter space of the problem by means of a series of simulations: the Reynolds number $Re_{\lambda}$ varies in the range $Re_{\lambda}\in [20, 150]$, the Schmidt number spans the interval $[0.1,4]$ and the Damkh\"oler number changes from $\mathcal{O}(10^{-4})$ to $\mathcal{O}(10)$, while \textcolor{black}{$n$ the reaction order for $R_2$} is increased from one up to $n=3$ (i.e. from second to fourth order \textcolor{black}{forward} reaction). The values of the key parameters for the simulations are reported in table~\ref{parameters}.

\begin{table}[ht]
\caption{\label{parameters}Parameters for the simulations: $Re_{\lambda}$ is the Taylor scale based Reynolds number; $Sc$ is the Schmidt number; $Da=\lambda \gamma_1 R_{2,eq}^n/u'=\lambda \gamma_2/u'$ is the Damkh\"{o}ler number based on Taylor scale; $n$ is the order of $R_2$ in the reaction; $N^3$ is the total number of grid points; $\eta$ is the Kolmogorov length or dissipative length; $|\mathbf{k}|_{max}$ is the maximum wave number amplitude kept by the dealiasing procedure; $|\mathbf{k}|_{max}\cdot\eta$ is the spatial resolution condition; $dt/\tau_{\eta}$ is the time step normalised by the Kolmogorov time scale $\tau_{\eta}$.}
\begin{ruledtabular}
\begin{tabular}{ccccc}
   No. & $1$  & $2$ & $3$ & $4$\\
  \hline
   $Re_\lambda$ & $20$  & $40$ & $80$ & $150$\\
   $Sc$ & $0.1$-$4$ & $0.1$-$4$ & $0.1$-$1$/$2$-$4$ & $1$ \\
   $Da$ & $0.0005$ - $50$ & $0.0003$ - $30$ & $0.0005$ - $20$ & $0.001$ - $10$ \\
   $n$ & $1$ & $1$ - $3$ & $1$ - $3$ & $1$ - $3$ \\
   $N^3$ & $64^3$ & $64^3$~\footnotemark[1]/$128^3$~\footnotemark[2] & $128^3$~\footnotemark[1]/$256^3$~\footnotemark[2] & $256^3$\\
   $|\mathbf{k}|_{max} \eta$ & $3.12$ & $1.48$~\footnotemark[1]/$2.95$~\footnotemark[2] & $1.26$~\footnotemark[1]/$2.52$~\footnotemark[2] & $1.05$ \\
   $dt/\tau_{\eta}$ & $0.044$ & $0.06$~\footnotemark[1] - $0.03$~\footnotemark[2] & $0.044$~\footnotemark[1] - $0.022$~\footnotemark[2] & $0.034$ \\
\end{tabular}
\end{ruledtabular}
\footnotetext[1]{corresponding to $Sc$ of $0.1$-$1$.}
\footnotetext[2]{corresponding to $Sc$ of $2$-$4$, better resolution condition is required to resolve the Batchelor micro scale ($\frac{\eta}{Sc^{1/2}}$).}
\end{table}

\section{Analysis of numerical results}\label{results}
\begin{figure}[ht]
\begin{center}
\subfigure[]{\includegraphics[width=0.5\linewidth]{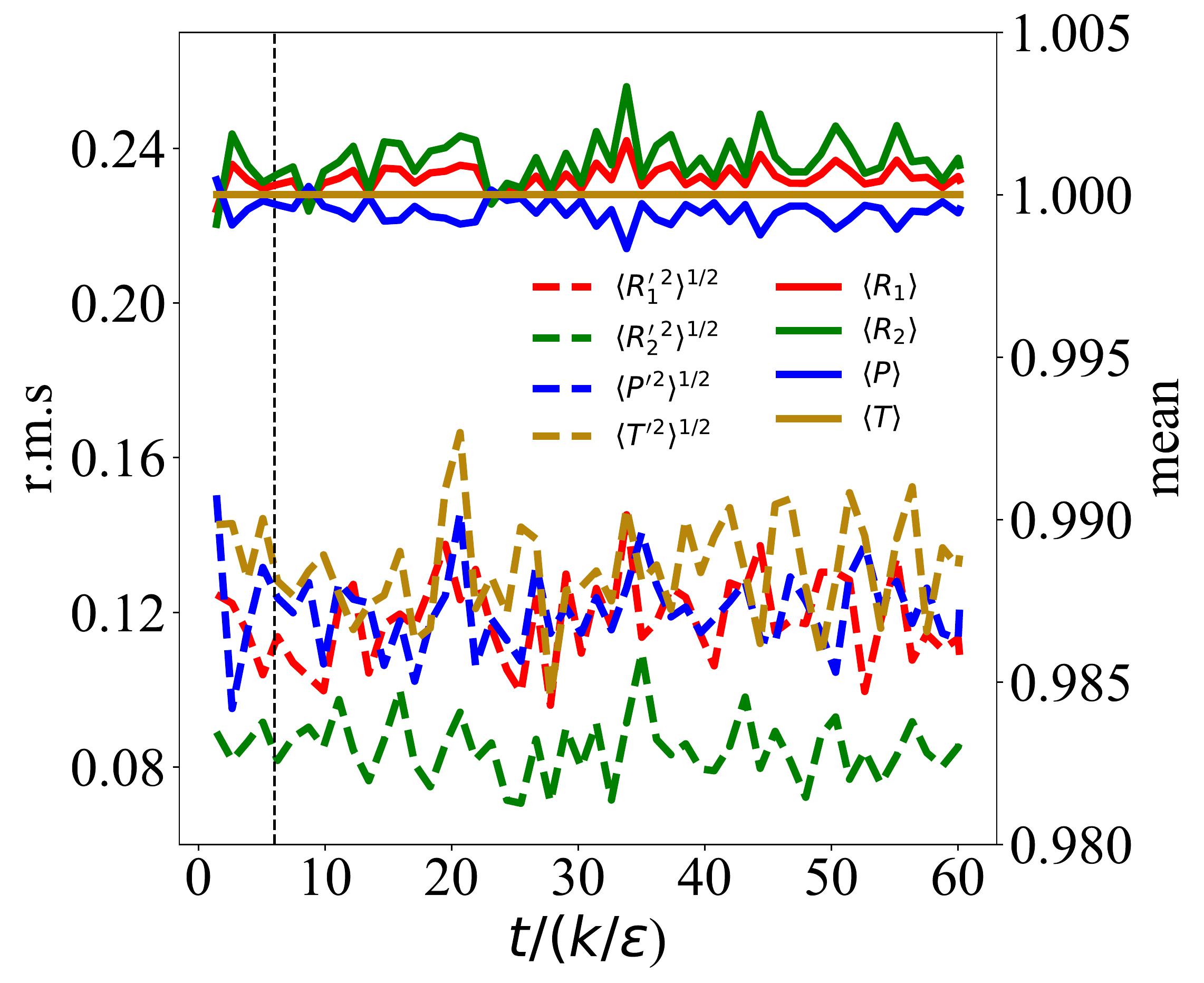}}
\subfigure[]{\includegraphics[width=0.425\linewidth]{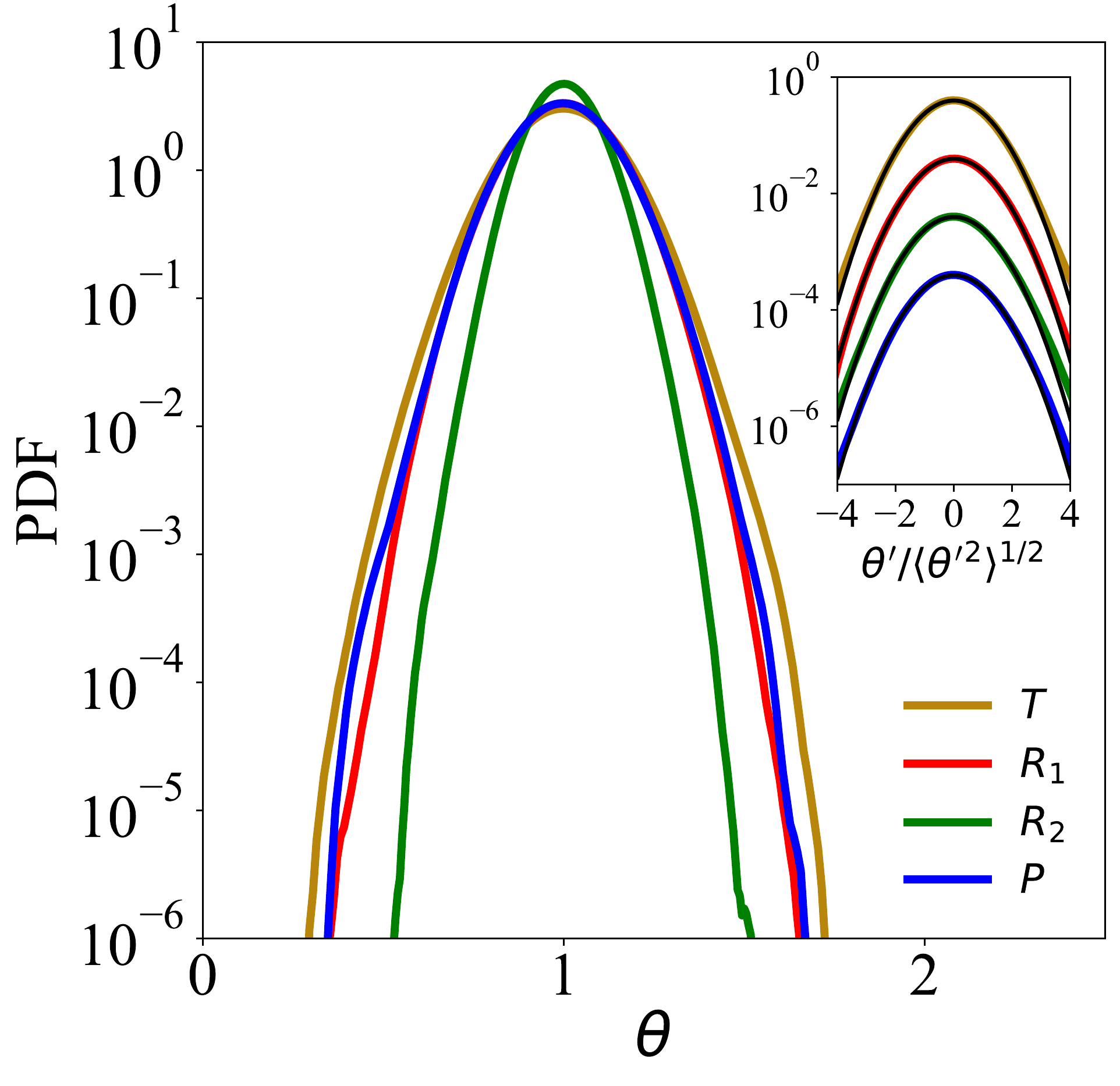}}
\end{center}
\caption{\label{evolution}a) Evolution of the root mean square of scalar fluctuations and mean values for the case of $n=2$, $Da=0.1$, $Sc=1$ and $Re_{\lambda}=150$. Time is normalised by the integral time $k/\varepsilon$ with $k=3u'^2/2$. The dash vertical line marks the initial time for the computation of statistical quantities in this study. b) PDF of scalar fields in dimensionless units (main panel) and normalised with respect to their standard deviations (inset, the black lines are Gaussian curves).}
\end{figure}

We begin by looking at the temporal evolution of the two first statistical moments of reactive fields, i.e., their mean values and root-mean-square fluctuations. The notation $\theta = \langle \theta \rangle_V + \theta'$ represents the decomposition into the mean $\langle \cdot\rangle_V$ (volume average) and fluctuation, where $\theta$ stands for a generic scalar quantity, either $R_1$ , $R_2$, $P$ or $T$,

Fig.~\ref{evolution} shows the typical temporal evolution of the fluctuating and the mean parts of scalars for a simulation with $n=2$, $Da=0.1$, $Sc=1$ and $Re_{\lambda}=150$. After a sufficiently long simulation time, a statistically steady state is established where the global mean value for the reactive scalar fields is close to the respective equilibrium quantities, \textit{i.e.}, $\langle R_1\rangle\approx\langle R_2\rangle\approx\langle P\rangle\approx1$. Furthermore, in spite of the presence of a vigorous external mechanical forcing and
random scalar source terms, the reactive scalar dynamics is characterised by relatively small global fluctuations from the equilibrium state.
We observe that the scalar fluctuations are proportional to the amplitude of the mechanical forcing, which poses a limitation for the numerical implementation of the model system, i.e. the positiveness of the scalar concentration fields $(\theta \geq 0)$.
In order to fulfil this constraint, in this study the r.m.s. of scalars reaches at maximum $10\%$ of the mean value.
The statistical convergence of our study is reached by means of simulations extending in time $\sim 45 T_{I}$, where $T_{I} = k/\varepsilon$ with $k=3u'^2/2$ is the integral time scale. The temporal averages are performed after at least $8 T_{I}$ from the beginning of the simulation (see Fig.~\ref{evolution}a).

The \textcolor{black}{corresponding} probability density functions (PDF) of the scalar fields are reported in Fig.~\ref{evolution}b. It can be seen that despite the different amplitudes of the standard deviations, their normalised shapes do not deviate significantly from Gaussian. No noticeable difference is observed in the comparison of reactive scalars with the passive one. \textcolor{black}{Furthermore, side-by-side visualisations of instantaneous snapshots of reactive and passive scalars do not allow to perceive any difference in their spatial structure.}
This lead us to argue that in the present conditions the differences between reactive and passive scalar are to be found in their global properties, such as correlation coefficients and global mean fluctuations, which will be the subject of the coming sections.

\textcolor{black}{We also note that in the above described statistically steady condition the advecting flow is the main responsible for local departures in time and space from the chemical equilibrium condition. The magnitude of such deviations depends on the turbulent intensity and it grows with $Re_{\lambda}$.
We observe that the root-mean-square intensity of the local reaction rate $Da( R_1 R_2^n -  P)$ goes approximately as $\sim Re_{\lambda}^{3/2}$ in the range of Reynolds number explored in this study.}

\subsection{Global correlation coefficients of reactive scalars}
In this section the global correlation coefficients for the scalar fields is investigated. We begin with a theoretical argument for the prediction of its functional dependence on varying the dimensionless \textit{a priori} control parameters $Re_{\lambda},Sc,Da$ and $n$. We will later compare the prediction with the numerical results.

The global correlation coefficients between reactive scalars are defined as
\begin{equation}
  r(\theta_1,\theta_2) = \frac{\langle \theta_1'\theta_2'\rangle}{\langle \theta_1'^2\rangle^{1/2}\langle \theta_2'^2\rangle^{1/2}}, \label{correlation_definition}
\end{equation}
\textcolor{black}{where} $\theta_1$ and $\theta_2$ are the \textcolor{black}{scalar fields} under consideration.

The theoretical prediction is based on the following two hypotheses. First, given the fact that the reactive scalar fluctuations are small with respect to the equilibrium global value, the chemical sources can be linearised in the following way:
\begin{equation}\label{Taylor_estimate}
    R_1R_2^n-P\approx(\langle R_1\rangle+R_1')(\langle R_2\rangle+n\langle R_2\rangle^{n-1}R_2')- (\langle P\rangle+P')\approx R_1'+nR_2'- P',
\end{equation}
where we have used also the fact that $\langle R_1\rangle\approx\langle R_2\rangle\approx\langle P\rangle\approx1$  (see again Fig.~\ref{evolution}a) ).
Second, we assume that there are no correlations between the source term for a given scalar and other reactive scalars, which is here reasonably guaranteed by the fact that the source terms are delta-correlated in time with a fixed amplitude. Such an assumption, however, is not a general feature of reactive turbulence, and needs to be considered specifically.

The global value of the cross product of scalar gradients $\langle {\bf \nabla} \theta_1 {\bf \nabla} \theta_2 \rangle$ can be estimated in terms of the global value of cross product of scalars $\langle  \theta_1  \theta_2 \rangle$ normalized by the square of a characteristic length-scale $\lambda_{\theta}$. For a scalar quantity $\theta$, we define $\lambda_{\theta}$ as in \citep{ristorcelli2006passive}:
\begin{equation}\label{lambda_theta}
  \lambda_{\theta}^2=\frac{\langle \theta'^2\rangle}{\langle (\partial_x \theta)^2\rangle}.
\end{equation}
Such a length can be interpreted as the Taylor micro-scale of $\theta$ (see also Appendix \ref{correlation_derivative}). Consequently, this allows to introduce an \textit{a posteriori} control parameter, the Damk\"{o}hler number based on the scalar Taylor micro-scale and diffusivity, denoted here $Da_{\theta}$, which is defined as
\begin{equation}\label{Da_theta}
  Da_{\theta}=Re_{\lambda} Sc Da \lambda_{\theta}^2=\frac{\lambda_{\theta}^2}{D}\gamma_1 R_{2,eq}^n.
\end{equation}
It has to be noted that such a number includes a combination of the three a priori control parameters ($Re_{\lambda}$,$Sc$,$Da$) for the model system, with the addition of the $\lambda_{\theta}$ scale, which therefore plays a key role in the analysis.

We provide here the central result of the derivation based on the above steps. The detailed derivation, lengthy but straightforward, is provided in the Appendix~\ref{theory_correlation}.
It is found that  $r(R_1,R_2)$, $r(R_1,P)$ and $r(R_2,P)$ are only dependent on the reaction order $n$ and $Da_{\theta}$. Specifically,
\begin{align}
  r(R_1,P)&\approx\frac{Da_{\theta}}{3+n^2Da_{\theta}+Da_{\theta}},\label{co_R1_P}\\
  r(R_1,R_2)=-r(R_2,P)&\approx\frac{-nDa_{\theta}}{\sqrt{3+n^2Da_{\theta}+Da_{\theta}}\sqrt{3+2Da_{\theta}}}.\label{co_R1_R2}
\end{align}
The above expressions show that the concentration field for $R_1$ is positively correlated to $P$, while is negatively correlated to $R_2$. Furthermore, the correlations $ r(R_1,R_2)$ and $r(R_2,P)$ are opposite in sign. They also show that for large $Da_{\theta}$ the correlations reach a saturation plateau, whose  value depends on the reaction order $n$.  For $n=1$ the correlations coefficients have all the same intensity and only differ in sign. The asymptotically large $Da_{\theta}$ limit in this case leads to the values $r = \pm 1/2$. At asymptotically large $n$ and $Da_{\theta}$, it yields $ r(R_1,P)\simeq 0$
and $r(R_1,R_2)=-r(R_2,P) \simeq 1$. On the opposite, in the condition of vanishing values of $Da_{\theta}$, corresponding to a negligible role of chemical processes and predominance of mixing, all the correlations coefficients tend to zero.

\begin{figure}[ht]
\centering
\includegraphics[width=0.95\linewidth]{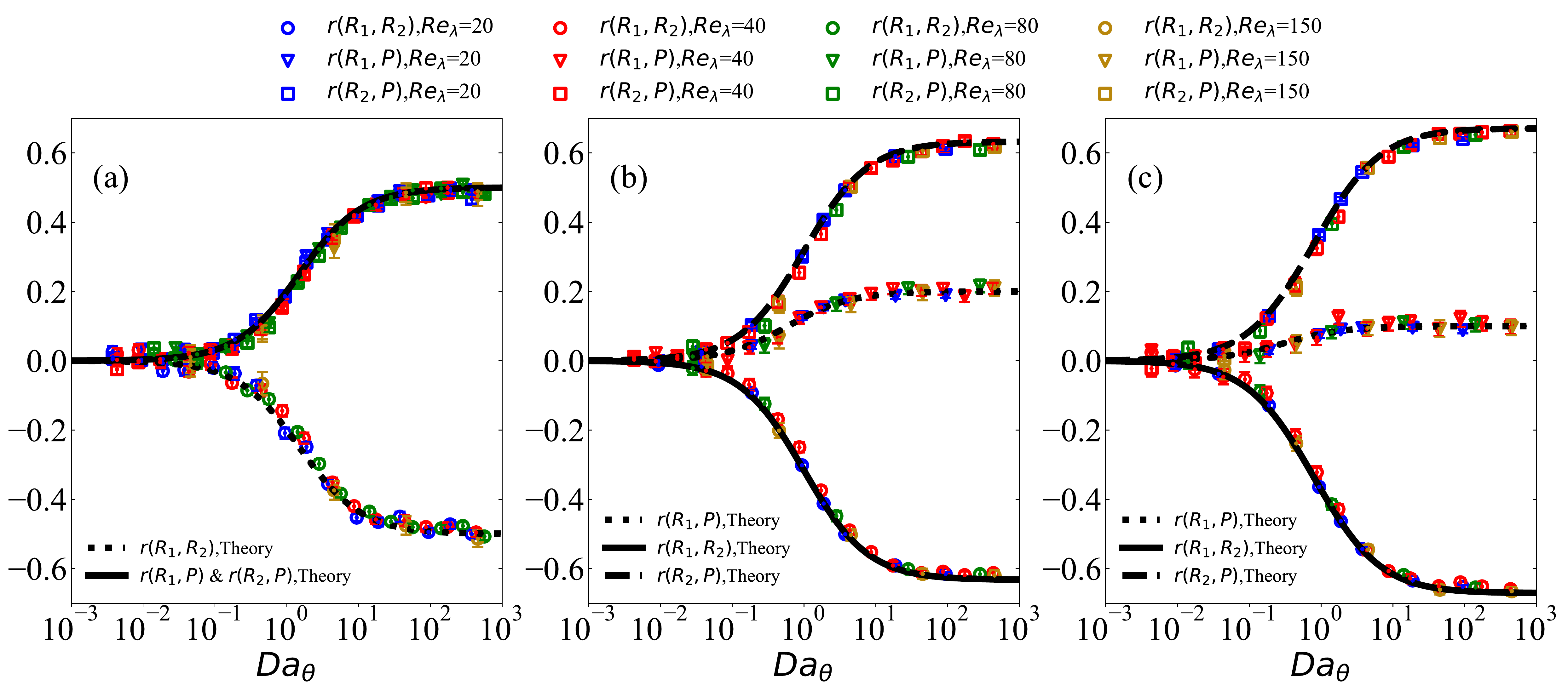}
\caption{Correlation coefficients between $R_1$ and $R_2$ ($r(R_1,R_2)$), $R_1$ and $P$ ($r(R_1,P)$), $R_2$ and $P$ ($r(R_2,P)$) as functions of $Da_{\theta}$, under the condition of (a): $n=1$ (b): $n=2$ and (c): $n=3$ and the Schmidt number $Sc=1$. Theoretical predictions are shown in black lines.}
\label{correlation_theta_Da}
\end{figure}

In Fig.~\ref{correlation_theta_Da} we report the numerical measurements of the correlation coefficients between the reactive scalars as functions of $Da_{\theta}$, for a set of simulations characterised by different reaction order $n$ and different $Re_{\lambda}$, ranging over more than a decade. It can be seen that for low $Da_{\theta}$ values, corresponding to slow reaction rates, $R_1$, $R_2$ and $P$ behave as almost independent passive scalars, as expected. Thus the correlation coefficients are about zero when $Da_{\theta}$ is small. As $Da_{\theta}$ increases, the scalars become more and more correlated and the correlation coefficients gradually approach plateaus, which are $n$ dependent. All these tendencies are in excellent agreement with the theoretical predictions, which are also reported on the same figure. 

\begin{figure}[ht]
\centering
\includegraphics[width=0.95\linewidth]{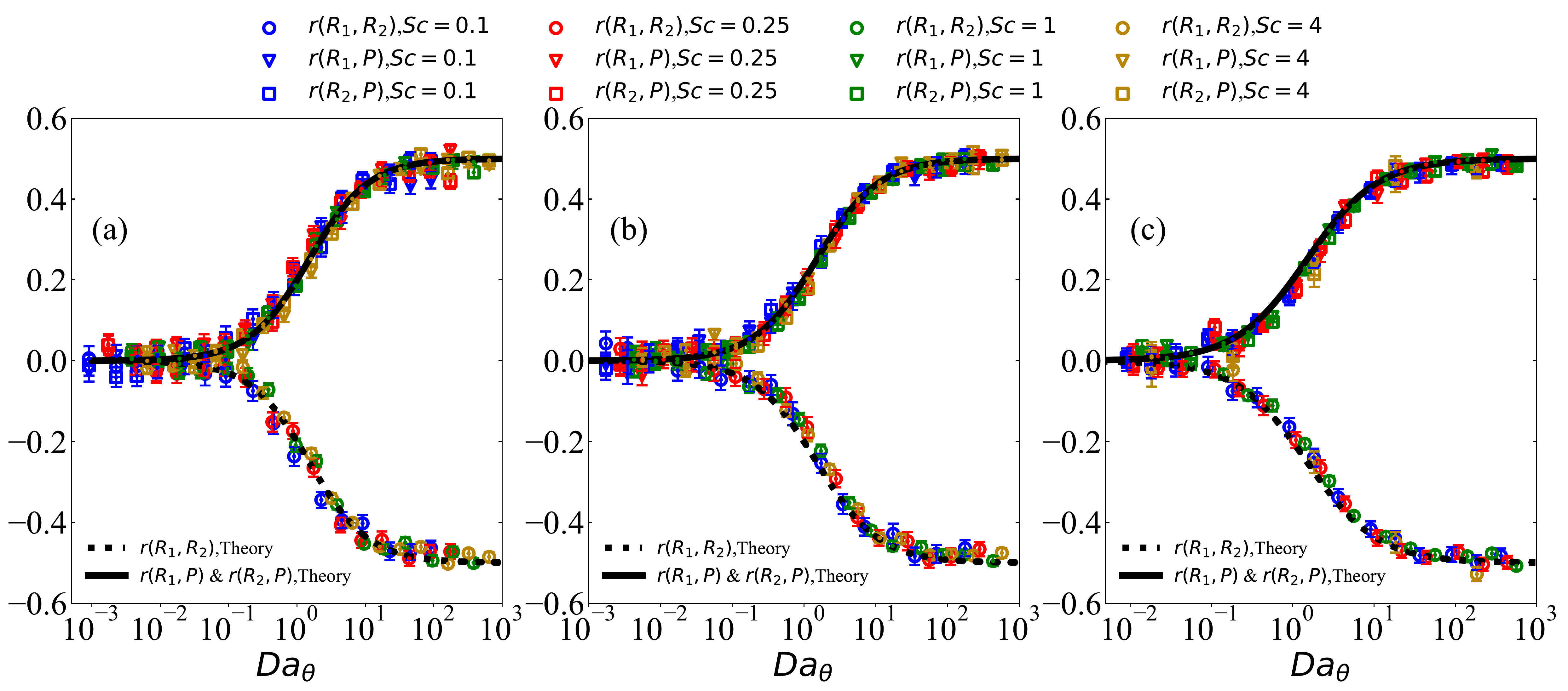}
\caption{Correlation coefficients between $R_1$ and $R_2$ ($r(R_1,R_2)$), $R_1$ and $P$ ($r(R_1,P)$), $R_2$ and $P$ ($r(R_2,P)$) as functions of $Da_{\theta}$, under the condition of (a): $Re_{\lambda}=20$ (b): $Re_{\lambda}=40$ and (c) $Re_{\lambda}=80$. The order of $R_2$ ($n$) is $1$. Theoretical predictions are shown in black lines.}
\label{correlation_theta_Da_different_Sc}
\end{figure}

Fig.~\ref{correlation_theta_Da_different_Sc} further confirms the range of validity of the prediction, by displaying the same correlations now for the case of different Schmidt numbers in the range from $0.1$ to $4$ and for $n=1$. Again the trends are well captured by the theoretical predictions.

\subsection{Variance of reactant scalar fields as compared to a passive scalar}

As we have already mentioned the turbulent advection and the scalar forcing are the sources of scalar spatial-temporal fluctuations.
In the case of passive scalars such fluctuations are smoothed out by diffusion. For the reactive case, the chemical sources function as an additional dumping mechanism. In other words, it is expected that the chemical reaction term acts as a global sink to suppress the scalar energy in addition to dissipation via molecular diffusion.

In the present model system this scenario can be understood  by means of the following argument. We multiply the linearized transport equations for the reactive scalars (equations \eqref{eq_scalar_non_dim_1_used_prime_M}-\eqref{eq_scalar_non_dim_3_used_prime_M} in appendix)
with the corresponding fluctuation field $R_1'$, $R_2'$ and $P'$ and perform volume and time average $\langle\cdot\rangle$. At statistical stationary state, summing the obtained equations of scalar dissipation rates ($\varepsilon_{R_1}$, $\varepsilon_{R_2}$ and $\varepsilon_{P}$) for the considered reactive fields reads:
\begin{align}
  \varepsilon_{R_1}+\varepsilon_{R_2}+\varepsilon_{P} & = \frac{\langle(\nabla R_1')^2\rangle}{Re_{\lambda}Sc}+\frac{\langle(\nabla R_2')^2\rangle}{Re_{\lambda}Sc}+\frac{\langle(\nabla P')^2\rangle}{Re_{\lambda}Sc}\nonumber\\
   & \approx - Da \langle R_1' +nR_2'-P'\rangle^2 + \langle R_1'\dot{q}_{R_1}\rangle + \langle R_2'\dot{q}_{R_2}\rangle + \langle P'\dot{q}_{P}\rangle.\label{total_dissipation}
\end{align}
The above equation shows that the reaction is always responsible of removing the scalar energy. A consequence of this is that one expects smaller scalar fluctuations for the reactive fields as compared to a passive scalar. In particular, we expect that the scalar variance will be a monotonically decreasing function in $Da_{\theta}$.

To have a quantitative understanding for such scenario we compare the fluctuations of the reactive scalars with the ones of a passive scalar in the same dynamical conditions, i.e., subject to the same advective flow, and having the same diffusion and under the effect of an independent statistical realisation of the source term $\dot{q}_{\theta}$.

In order to develop also in this case a quantitative prediction for the phenomenon we need to introduce the key assumption that the scalar energy input due to the source term on the field $R_1$ is approximately same as the one provided on a passive scalar field $T$ in the same conditions, i.e.
\begin{equation}
\langle R_1' \dot{q}_{R_1}\rangle \simeq \langle T' \dot{q}_{T}\rangle.
\end{equation}
The soundness of this hypothesis lies on the fact that in the present conditions the reactant $R_1$ has fluctuation of similar intensity as the passive scalar case.

Following a similar derivation as for Eq.~\eqref{co_R1_P} and~\eqref{co_R1_R2}, the fluctuations of the reactive scalars normalized by the fluctuation of passive scalar ($T$) are estimated as
\begin{align}
  \frac{\langle R_2'^2\rangle}{\langle T'^2\rangle} \approx& \frac{3+2Da_{\theta}}{3+n^2Da_{\theta}+2Da_{\theta}},\label{R2_over_T}\\
  \frac{\langle R_1'^2\rangle}{\langle T'^2\rangle}=\frac{\langle P'^2\rangle}{\langle T'^2\rangle} \approx& \frac{3+n^2Da_{\theta}+Da_{\theta}}{3+n^2Da_{\theta}+2Da_{\theta}}.\label{R1_P_over_T}
\end{align}
For the full details for this derivation, which implies performing spatial and temporal averaging over the scalar energy equations, the reader is referred to Appendix~\ref{theory_fluctuations}. It is found that the fluctuations of the reactive scalars ($R_1$, $R_2$ and $P$) are close to that of passive scalar ($T$) when $Da_{\theta}$ is small, but gradually decrease as $Da_{\theta}$ increases.
\begin{figure}[ht]
\centering
\includegraphics[width=0.95\linewidth]{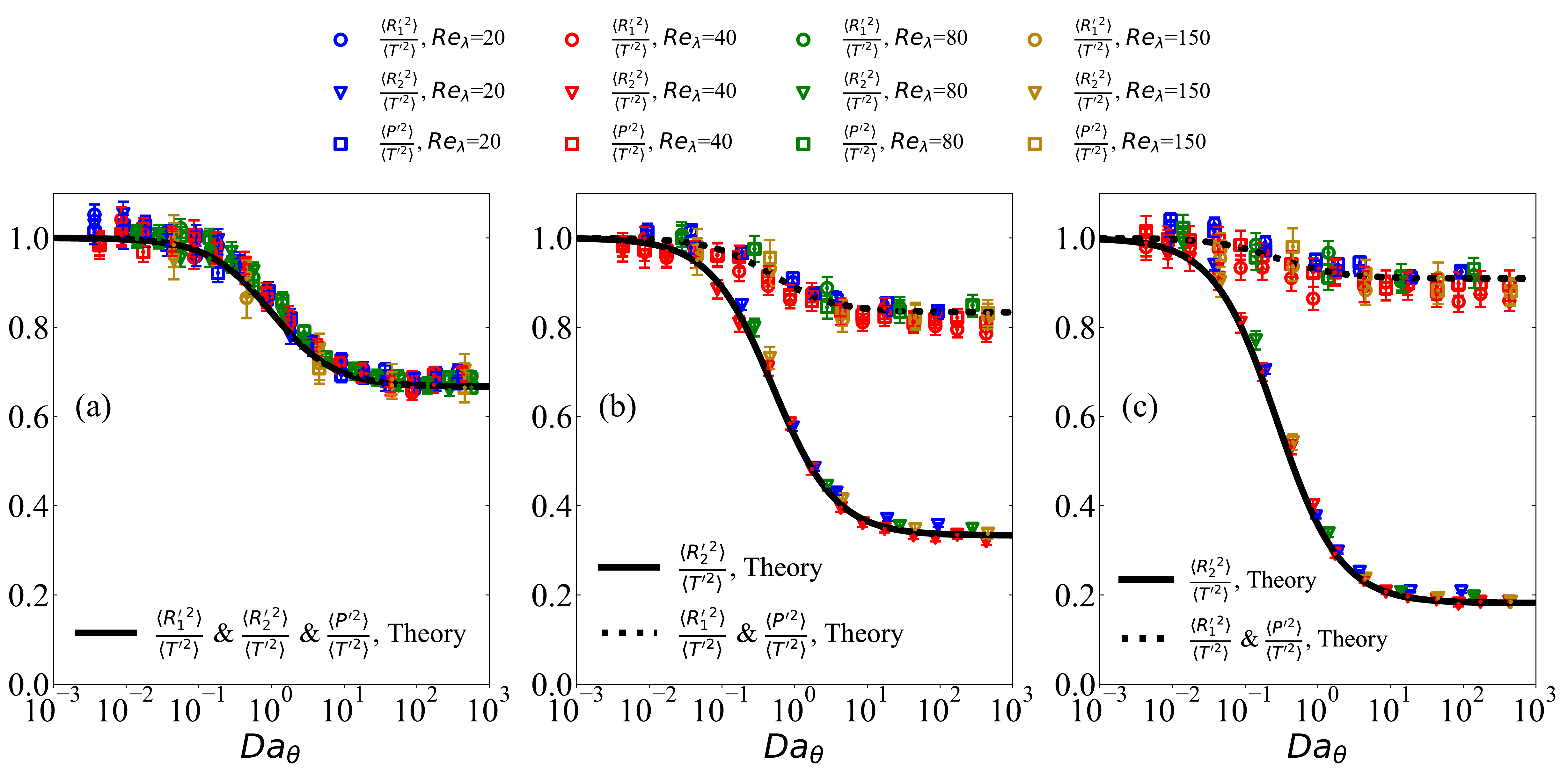}
\caption{The fluctuations of the reactive scalars normalised by the fluctuation of passive scalar ($T$) as functions of $Da_{\theta}$, under the condition of (a): $n=1$ (b): $n=2$ and (c): $n=3$ and the Schmidt number $Sc=1$. Theoretical predictions are shown in black lines.}
\label{nor_rms_theta_Da}
\end{figure}
Figure~\ref{nor_rms_theta_Da} shows the normalized fluctuations of the reactive scalars ($R_1$, $R_2$ and $P$) as measured from the DNS, in agreement with the above theoretical prediction.

\subsection{On the Taylor micro scale of scalar concentration fields}
As we have discussed in the above sections the correlations and fluctuations of the concentration of reactive scalars are well described by means of the control parameter $Da_{\theta}$, which contains the scalar Taylor micro scale of $\lambda_{\theta}$. In this section we aim at gaining more insight into this key spatial scale.

The scalar Taylor micro-scale was notably first studied by S. Corrsin \cite{corrsin1957simple}, in the context of scalar mixing in turbulent flows. He hypothesized that such a scale is proportional to the intensity of turbulence and inversely to the Schmidt number of the problem, i.e. \[\frac{\lambda_{\theta}^2}{\lambda^2}\propto\frac{1}{Sc}.\] However, further experimental and numerical studies  \citep{bahri2015self,corrsin1964isotropic} have reported that such a dependence is not straightforward, as it shows finite $Re_{\lambda}$ effects and different trends for asymptotically small and large $Sc$ values (see \citep{ristorcelli2006passive} for a recent discussion).

\begin{figure}[ht]
\begin{center}
\subfigure[]{\includegraphics[width=0.45\linewidth]{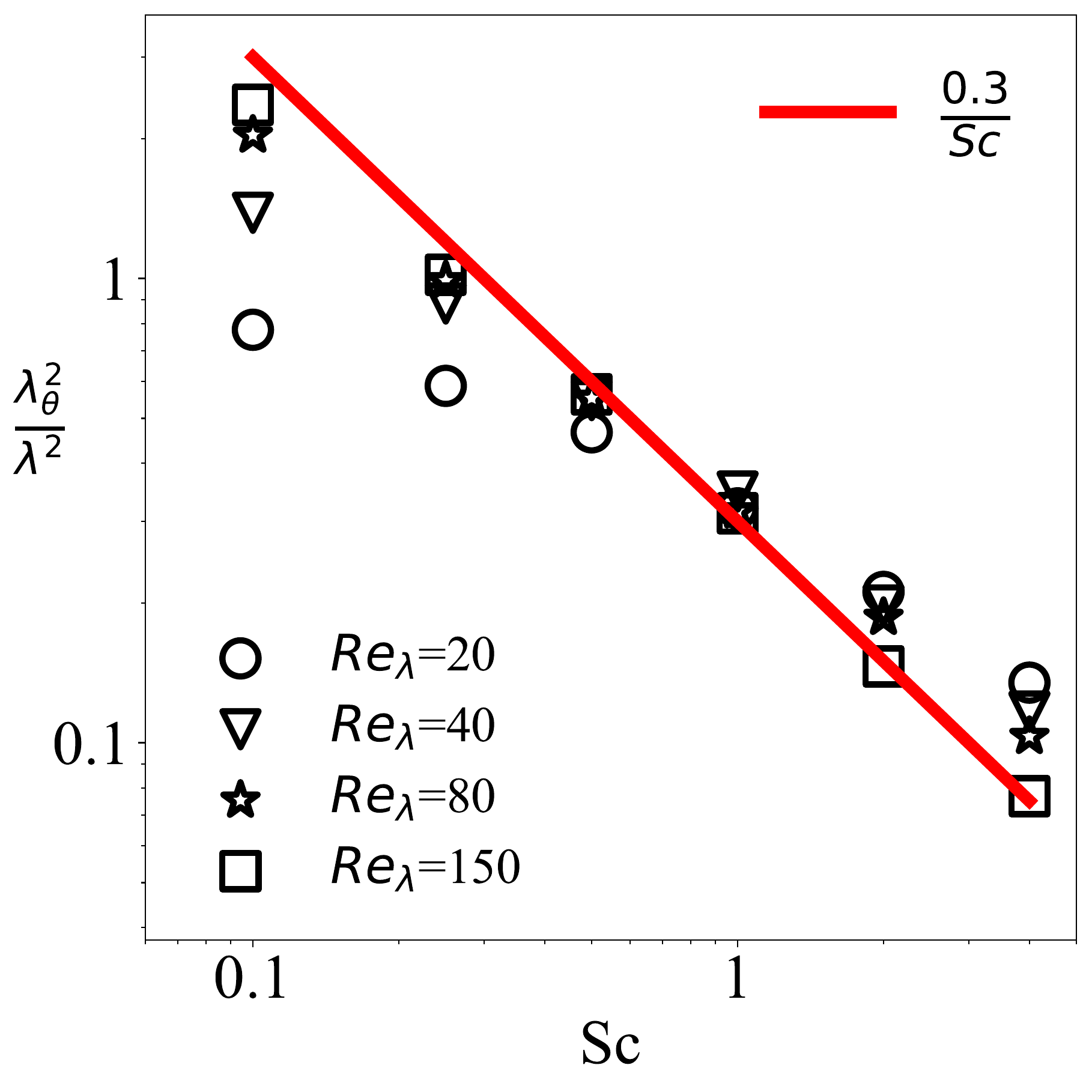}}
\subfigure[]{\includegraphics[width=0.45\linewidth]{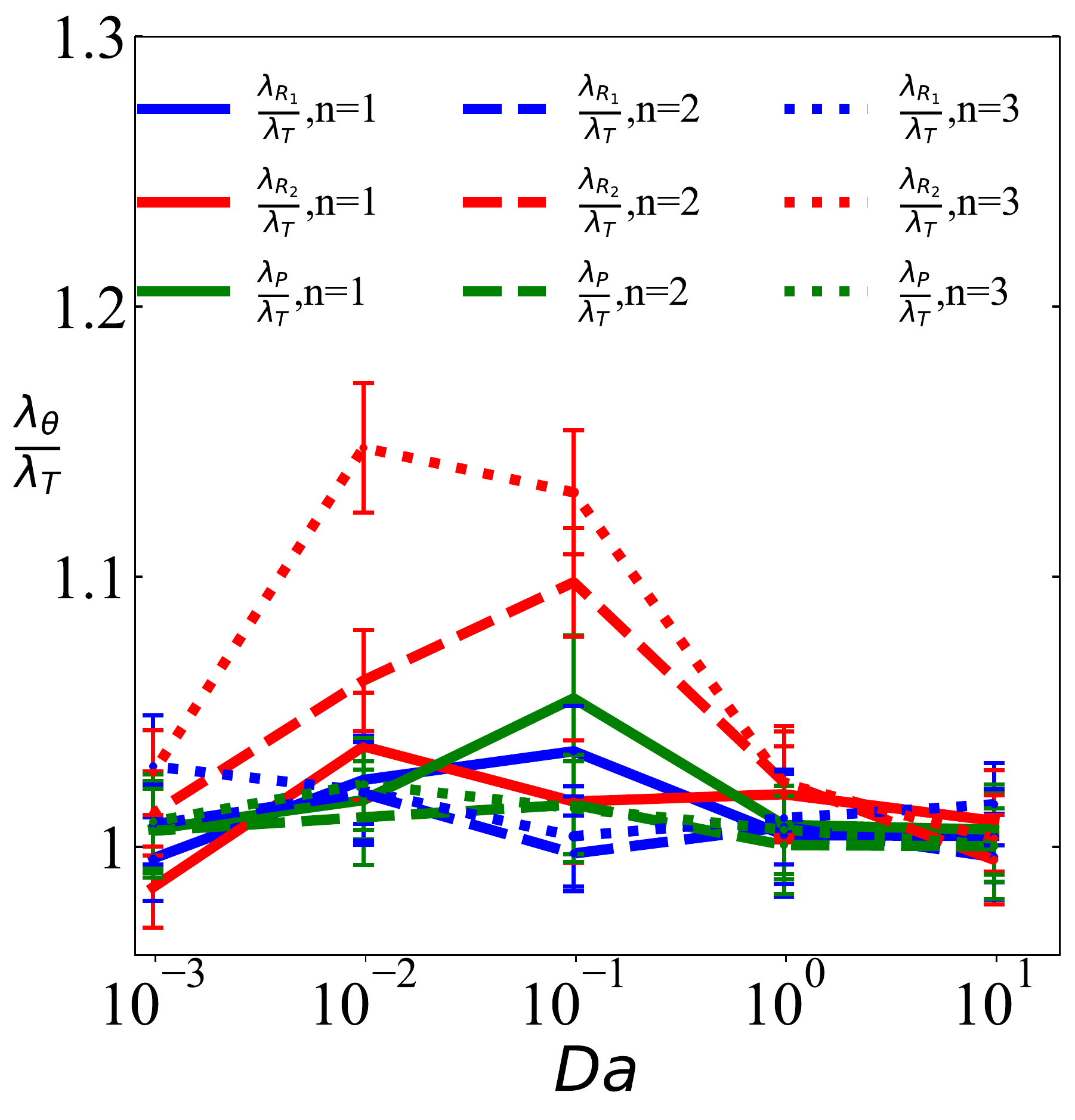}}
\end{center}
\caption{ a) Taylor micro scales of scalars (computed on $T$) as functions of $Sc$. The red line draw is $0.3/St$. A power law fit of the form $a Sc^b$ on the $Re_{\lambda}=150$ data set gives $a=0.29 \pm 0.1$ and $b= -0.93 \pm 0.02$. \textcolor{black}{b) Taylor micro scales of the reactive scalars with respect to  the passive scalar one with different order of reaction, as functions of $Da$ for all the simulations at $Re_{\lambda}=150$ and $Sc=1$.}}
\label{lambda_sq_q_over_lambda_sq_u}
\end{figure}

We show the results of our simulations in Fig.~\ref{lambda_sq_q_over_lambda_sq_u}a. The figure  reports the dimensionless $\lambda_{\theta}^2$ (actually $\lambda_{\theta}^2 / \lambda^2$, because the length is non-dimensionalised by $\lambda$) as functions of $Sc$ under the conditions of $Re_{\lambda}$ from $20$ to $150$. Fig.~\ref{lambda_sq_q_over_lambda_sq_u} indicates that a clear $-1$ scaling law exists between $\lambda_{\theta}^2$ and $Sc$ for the largest $Re_{\lambda}$ case, with a prefactor  $\simeq 0.3$, implying that in the limit of intense turbulence the $Da_{\theta}$ number can be approximated as
\begin{equation}
Da_{\theta} \simeq 0.3 Re_{\lambda} Da = 3 T_{I} / \tau_{r}.
\end{equation}
Here $\tau_{r} = (\gamma_1 R_{2,eq})^{-1}$ is the typical time of the reaction.
Remarkably, this result reveals that the unique \textit{a posteriori} control parameter that we have identified with $Da_{\theta}$ can be considered as the ratio of the largest time scale of the turbulent flow to the typical time scale associated to the chemical process.

\textcolor{black}{Finally, we remark that $\lambda_{\theta}$ does not vary significantly over the different scalar fields $R_1$, $R_2$, $P$ and the reference passive scalar field $T$. This is exemplified in Figure  \ref{lambda_sq_q_over_lambda_sq_u} for all the simulations at $Re_{\lambda}=150$ and $Sc =1$.  The figure shows that any $\lambda_{\theta}$ evaluated on a reactive field is at best $15\%$ different from the reference $\lambda_T$ case. Such difference vanishes for very small (the mixing dominated limit) or very large $Da$ and shows a weak increase trend with the order of the reaction. We can conclude that $\lambda_T$ can here be taken as a convenient approximation of $\lambda_{\theta}$. The estimations of $Da_{\theta}$  in the present work are based on such an assumption. 
}

\subsubsection{Taylor micro scale of scalars advected by a coarse-grained turbulent flow field}

In order to understand better the role of $\lambda_{\theta}$, we perform a series of simulations where the scalar fields are advected by a coarse-grained, \textit{i.e.} spatially filtered turbulent flow, denoted as $\tilde{\mathbf{u}}$. The filter is a spectral low pass, defined as
\begin{equation}
 \hat{\tilde{\mathbf{u}}}=\sum_{|\mathbf{k}|\leqslant K}\hat{\mathbf{u}}(\mathbf{k}),
\end{equation}
where $K$ specifies the  maximum wave number kept in the modified field. Such a filter retains only the large eddies of the turbulent flow, down to a wavelength
$2 \pi / K$.
The Taylor scale for scalars convected by the filtered flow is denoted as $\tilde{\lambda}_{\theta}$. It is noteworthy that the length quantities are always adimensionalised by the Taylor scale of unfiltered flow $\lambda$, instead of the Taylor scale of the filtered flow. It is worth exploring what is the impact of the hierarchy of flow scales, extending from the domain size down to the dissipative scales, on the reactive scalar dynamics. In particular, we aim at understanding the dependence of the scalar correlation coefficients and the scalar Taylor micro-scale on the maximum wavenumber $K$. According to the above discussion, it is reasonably expected that the small scales of the fluid have a negligible influence, because the scaling mixing process and relevant quantities are controlled by the large eddy turnover time $T_I$.

\begin{figure}[ht]
\centering
\includegraphics[width=0.5\linewidth]{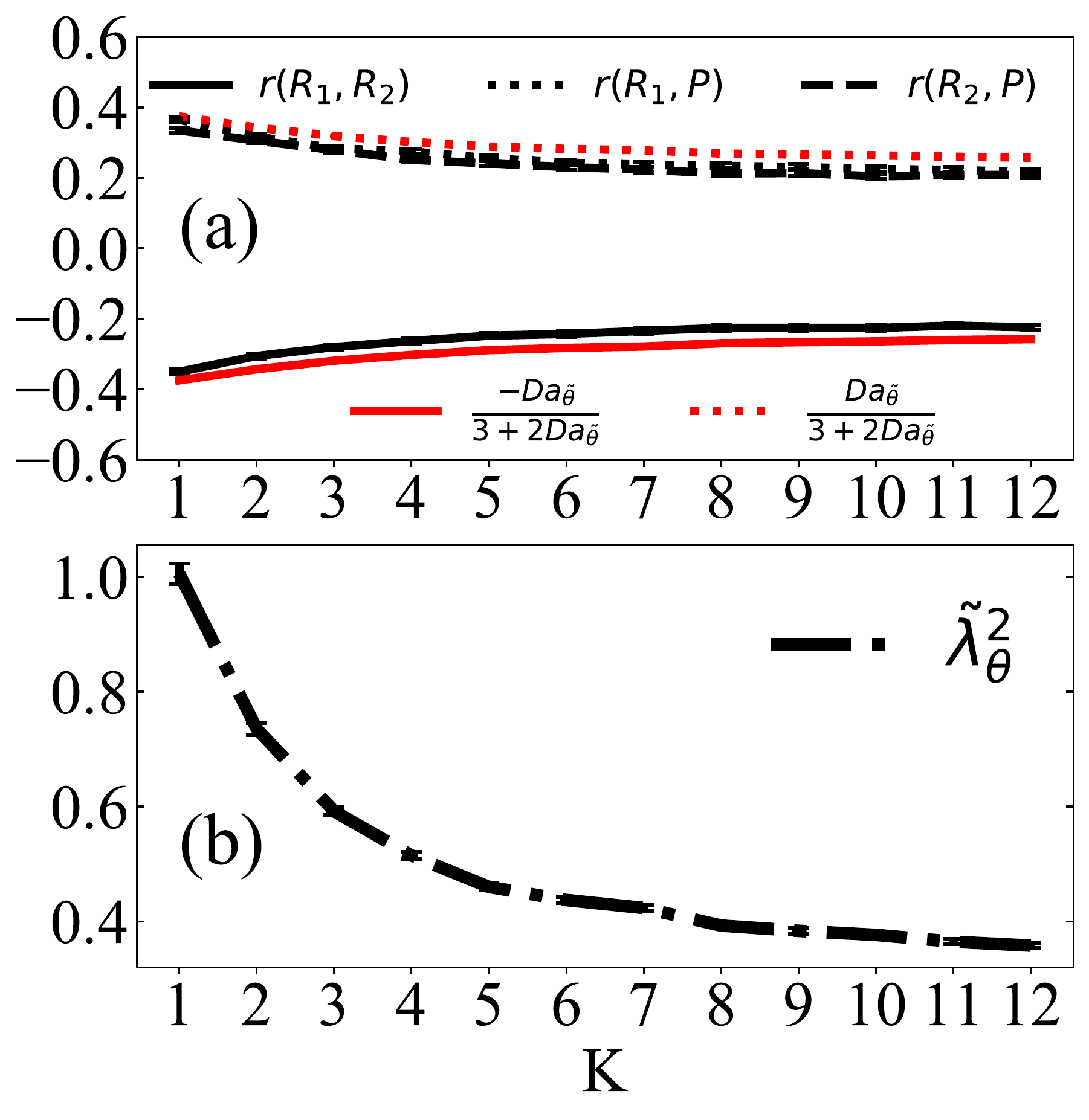}
\caption{(a) Correlation coefficients of scalar evolving in a coarse-grained turbulent flow filed, compared with theoretical predictions and (b) Taylor micro scales of the scalars convected by filtered velocity ($\tilde{\lambda}_{\theta}$) as functions of the maximum wave number of filtered velocity ($K$), at $Sc=1$, $Re_{\lambda}=80$ and $Da=0.05$ ($Da_{\theta}=1.42$).}
\label{ps_correlation_lambda_filter_u_n_1}
\end{figure}

Results are presented for a typical case, i.e. $Sc=1$, $Re_{\lambda}=80$ and $Da=0.05$ corresponding to  $Da_{\theta}=1.42$. Fig.~\ref{ps_correlation_lambda_filter_u_n_1}(a) shows that the correlation coefficients vary quite weakly with the filter parameter $K$. The deviation becomes noticeable only for $K\leq 3$, which corresponds to scales larger than the large eddy turnover scale of the flow, in the sense that the forcing is active up to $|\textbf{k}|=2\sqrt{2}$. Such behaviour is also well captured by the theoretical predictions of (\ref{co_R1_P}),(\ref{co_R1_R2}) if the Damkholer number $Da_{\tilde{\theta}}$ adopted is built on the measured scalar Taylor scale $\tilde{\lambda}_{\theta}$, instead of the original $Da_{\theta}$. This confirms again the relevance of the
scalar Taylor micro-scale in characterising the present reactive scalar system. Finally we note that, differently from the correlation coefficients, the scale $\tilde{\lambda}_{\theta}$ varies sensibly with the filter wavenumber $K$, as reported in  Fig.~\ref{ps_correlation_lambda_filter_u_n_1}(b).

\subsection{Spectra and coherency spectra of reactive scalars}
In this section we focus on the scale-dependent behaviour of reactive scalar fluctuations and their mutual correlation.
\subsubsection{Energy spectrum}

\begin{figure}[ht]
\centering
\includegraphics[width=0.5\linewidth]{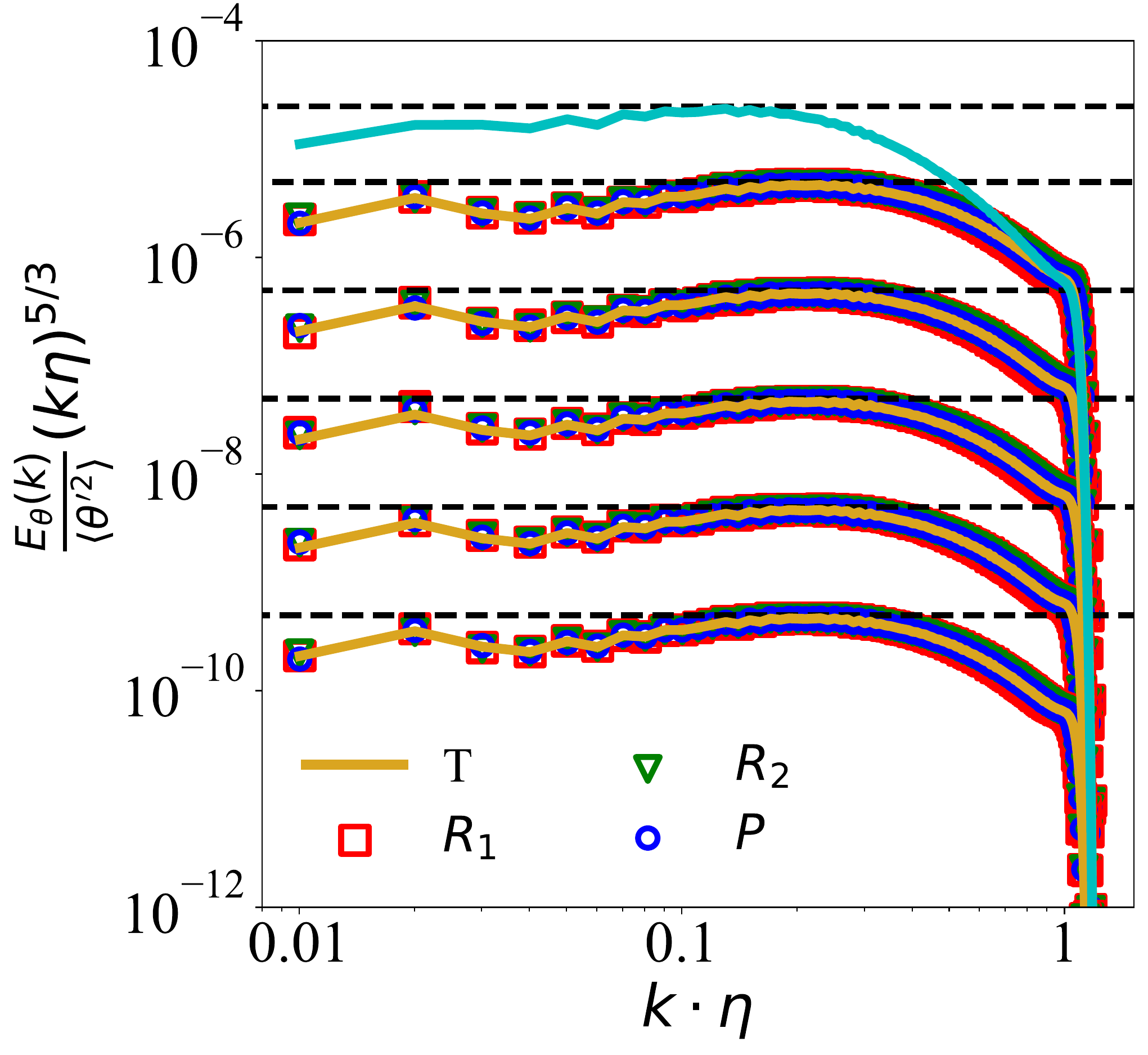}
\caption{
Energy spectra of reactive and passive scalar fields, i.e. $E_{\theta}(k)$ with $\theta=R_1, R_2, P, T$, and velocity field $E_u(k)$ (solid light-blue line) in the condition $Re_{\lambda}=150$, $Sc=1$, $n=1$ and $Da = 10, 1, 0.1, 0.01, 0.001$ (from top to bottom). Each spectra is compensated with the KOC scaling, $(k\eta)^{5/3}$ and normalised by the global energy $\langle \theta'^2 \rangle$. For clarity, the energy spectra of scalars are shifted vertically by a multiplicative factor 0.1}
\label{ps_spec_n_is_1}
\end{figure}
The energy spectra of the velocity and scalars are defined as
\begin{align}\label{energy_spectra}
  E_{u}(k)=& 4\pi k^2\langle \frac{1}{2}\hat{u}_i(\mathbf{k})\hat{u}_i^*(\mathbf{k})\rangle_k,\\
  E_{\theta}(k)=& 4\pi k^2\langle \hat{\theta}(\mathbf{k})\hat{\theta}^*(\mathbf{k})\rangle_k, \quad \theta=R_1, R_2, P \text{ and } T,
\end{align}
where $\langle\cdot\rangle_k$ denotes the average over all the modes in the shell of thickness $\Delta k$ centred at $k = |\mathbf{k}|$ and time, $\hat{u}_i(\mathbf{k})$ and $\hat{\theta}(\mathbf{k})$ are the Fourier coefficients of the mode of $\mathbf{k}$, $\hat{u}_i^*(\mathbf{k})$ and $\hat{\theta}^*(\mathbf{k})$ are the corresponding complex conjugate fields.

Figure ~\ref{ps_spec_n_is_1} depicts the log-log plots of the three-dimensional energy spectra of scalars of a typical case at $Re_{\lambda}=150$, $Sc=1$, $n=1$, compensated with $\sim k^{-5/3}$, which is the scaling expected in the inertial regime both for the velocity and for a passive scalar field, i.e., the KOC scaling \cite{Kolmogorov1941a,Kolmogorov1941b}. The spectra are also normalised by the total energy for each scalar. The figure shows that the spectra are indistinguishable from the ones of a passive scalar, and display the same scaling in the inertial range. This proves that in the present condition the reaction terms have a negligible effect on the scalar energy transfer.
Remarkably, this behaviour is also Damkh\"{o}ler  number independent. \textcolor{black}{The latter observation is qualitatively confirmed also by visualisations of the instantaneous scalar fields for different $Da$ values. Despite perceptible larger fluctuations for the small $Da$ case (the one where the chemistry is slower) it appears that the spatial structure of the fields is not affected by the magnitude of $Da$.}

\subsubsection{Coherency spectra}
The coherency spectrum between two scalar fields  $\theta_1$ and $\theta_2$ is defined as
\begin{equation}\label{cospectra_scalar}
  Co_{\theta_1,\theta_2}(k)=\frac{\langle|\widehat{\theta_1}(\mathbf{k})\widehat{\theta_2}^*(\mathbf{k})|\rangle_k}{\sqrt{\langle\widehat{\theta_1}(\mathbf{k})\widehat{\theta_1}^*(\mathbf{k})\rangle_k\langle\widehat{\theta_2}(\mathbf{k})\widehat{\theta_2}^*(\mathbf{k})\rangle_k}}.
\end{equation}
This function describes the scale dependence, in spectral space, of the correlation coefficient between two scalar fields.
\begin{figure}[ht]
\centering
\includegraphics[width=0.5\linewidth]{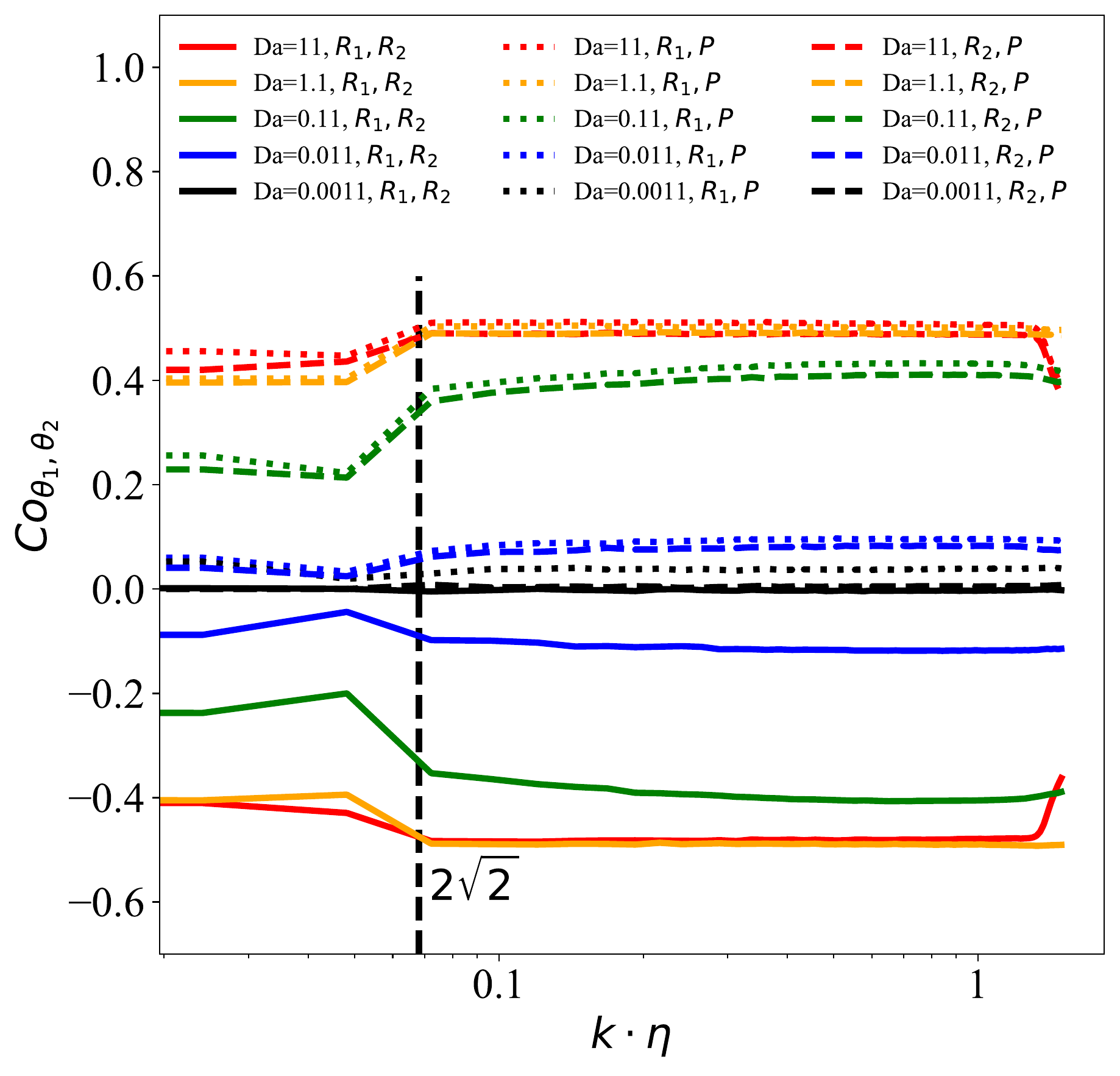}
\caption{Coherency  spectra of the reactive scalars, under the condition of $Re_{\lambda}=80$, $Sc=1$, $n=1$. The horizontal axis is normalised in terms of the Kolmogorov scale $\eta$. The dash vertical line marks the maximum wave number at which the scalar source terms acts.}
\label{cospec_n_is_1}
\end{figure}
Figure ~\ref{cospec_n_is_1} shows the results for a typical case at fixed $Re_{\lambda}=80$, $Sc=1$, $n=1$ and varying $Da$. It is interesting to note that for all the $Da$ cases the coherency spectra are nearly $k$ independent; in particular, they are constant in the inertial range.
A non constant behaviour is observed at small wave-numbers \textcolor{black}{$|\textbf{k}| \leq 2\sqrt{2}$}, which correspond to the largest physical scales. This is due to  the action of the random scalar source that strongly reduces the intensity of correlations.
Moreover, the absolute value of $Co_{\theta_1,\theta_2}$ increases as $Da$ increases, which agrees with the picture that fast chemical reactions build up correlations.


\section{Conclusions and Outlooks}\label{conclusion}
In summary, the statistical properties of species undergoing reversible chemical reactions in a turbulent environment have been studied. We have addressed this by means of a model system in which the flow is statistically steady turbulence and the chemical species are retained in a dynamical equilibrium state due to the action of random large-scale source terms. It is observed that the reactive scalar fluctuations have a Gaussian distribution and energy spectra are essentially identical to the one of a passive scalar field transported by the same flow. This can be explained by the overall small amplitude of the reaction terms in the present close-to-equilibrium conditions.

However, in such a state a competition still exists between the chemical processes, which tend to dump reactant concentration fluctuations and enhance their correlation intensity, and the turbulent mixing, which on the contrary increases fluctuations and removes relative correlations.

We quantitatively describe this phenomenon by considering the linearised equations for the reactive scalar fluctuations. A unique control parameter, the Damkh\"{o}ler number ($Da_\theta$), can be constructed as the ratio between the time scale of scalar diffusion across a domain of the size of the scalar Taylor micro-scale ($\lambda_{\theta}^2/D$) and the chemical reaction time scale $\tau_r$. Importantly, $Da_\theta$ characterises the functional dependence of fluctuations and correlations of the scalar quantities in the full range of explored conditions with variable reaction order, the Reynolds number and the Schmidt number. The larger is such a Damkh\"{o}ler number the more depleted are the scalar fluctuations as compared to the fluctuation of a passive scalar field in the same conditions, and vice-versa the more intense are the correlations. A saturation in this behaviour is observed beyond $Da_\theta \simeq \mathcal{O}(10)$.  This results reveal the significance of the scalar Taylor micro-scale for problems involving the mixing of chemical species.  We have shown that in the limit of intense turbulence the relation proposed by Corrsin \citep{corrsin1957simple} $\lambda_{\theta} \sim \lambda Sc^{-1/2}$ holds approximately, meaning that $Da_\theta$ can also be viewed as the ratio of the large-eddy-turnover time of the flow over the typical chemical reaction time.

The latter observation can be of practical interest for the estimation of the regimes attained by biogeochemical reactions at small scales in the ocean. For example, if one considers  a reactive field as the local concentration of phytoplankton in the ocean. The times scales used to construct the $Da_{\theta}$ number are: the typical large eddy-turnover time of the three-dimensional turbulent flow, which is normally of the order of hours \citep{Jimenez1997}; the typical growth rate of the population that is of the order of a day. This leads to value of  $Da_{\theta}<1$, meaning that, at small scales, more precisely at scales where the oceanic flow can be approximated as a three-dimensional turbulent flow with a direct energy cascade, the phytoplankton concentration can be safely considered as a passive scalar field.
However, differently from the situation explored in this work, the fluctuations displayed by planktonic populations can attain values that are comparable to the ones of the mean population density. This means that there exist regions where nearly any individual can be found or where huge accumulation can be observed. These situations go beyond the model explored in this work, and beyond the linearised theoretical predictions that we put forward. For this reason, it will be interesting to explore in the future different flow configurations where strong chemical sources could produce \textcolor{black}{pronounced deviations} from the global equilibrium state.

\appendix
\section*{APPENDIX: Analytical derivations}
In this Appendix we provide the detailed analytical arguments for the relation between the correlation spectrum of reactive scalar fields and their gradient spectrum,  as well as the analytical derivations for the prediction of global reactive scalar correlations and variances.

\section{Analytical prediction for reactant correlations}\label{theory_correlation}
Under the conditions of $R_1'\ll\langle R_1\rangle$, $R_2'\ll\langle R_2\rangle$, $P'\ll\langle P\rangle$ and $\langle R_1\rangle\approx\langle R_2\rangle\approx\langle P\rangle\approx1$ (Fig.~\ref{evolution}), the net reaction rate can be estimated as
\begin{equation}
    R_1R_2^n-P\approx(\langle R_1\rangle+R_1')(\langle R_2\rangle+n\langle R_2\rangle^{n-1}R_2')- (\langle P\rangle+P')\approx R_1'+nR_2'- P'.
\end{equation}
Correspondingly, the equations for fluctuating scalars become
\begin{align}
  \label{eq_scalar_non_dim_1_used_prime_M}\frac{D R_1'}{D t}\approx&\frac{1}{Re_{\lambda}Sc}\triangle R_1'-Da (R_1'+nR_2'+M')+\dot{q}_{R_1},\\
  \label{eq_scalar_non_dim_2_used_prime_M}\frac{D R_2'}{D t}\approx&\frac{1}{Re_{\lambda}Sc}\triangle R_2'-nDa (R_1'+nR_2'+M')+\dot{q}_{R_2},\\
  \label{eq_scalar_non_dim_3_used_prime_M}\frac{D M'}{D t}\approx&\frac{1}{Re_{\lambda}Sc}\triangle M'-Da (R_1'+nR_2'+M')+\dot{q}_M,
\end{align}
where $M'=-P'$ and $\dot{q}_M=-\dot{q}_P$. The same form of Eq.~\eqref{eq_scalar_non_dim_3_used_prime_M} as Eq.~\eqref{eq_scalar_non_dim_1_used_prime_M} implies that $R_1'$ and $M'$ behave statistically the same.

By multiplying Eq.~\eqref{eq_scalar_non_dim_1_used_prime_M},~\eqref{eq_scalar_non_dim_2_used_prime_M} and~\eqref{eq_scalar_non_dim_3_used_prime_M} with $R_2'$, $M'$ and $R_1'$ respectively and averaging ($\langle \dots \rangle$) on time and space, it yields:
\begin{align}
  \langle\frac{D R_1'}{D t}R_2'\rangle+\frac{1}{Re_{\lambda}Sc}\langle\nabla R_2'\nabla R_1'\rangle&\approx-Da (\langle R_2'R_1'\rangle+n\langle R_2'^2\rangle+\langle R_2'M'\rangle)+\langle R_2'\dot{q}_{R_1}\rangle,\label{eq_R1_R2}\\
  \langle\frac{D R_2'}{D t}M'\rangle+\frac{1}{Re_{\lambda}Sc}\langle\nabla M'\nabla R_2'\rangle&\approx-nDa (\langle M'R_1'\rangle+n\langle M'R_2'\rangle+\langle M'^2\rangle)+\langle M'\dot{q}_{R_2}\rangle,\label{eq_M_R2}\\
  \langle\frac{D M'}{D t}R_1'\rangle+\frac{1}{Re_{\lambda}Sc}\langle\nabla R_1'\nabla M'\rangle&\approx-Da (\langle R_1'^2\rangle+n\langle R_1'R_2'\rangle+\langle R_1'M'\rangle)+\langle R_1'\dot{q}_{M}\rangle.\label{eq_R1_M}
\end{align}

In Eq.~\eqref{eq_R1_R2}, the term $\langle R_2'\dot{q}_{R_1}\rangle$ is estimated as $0$, because the time-delta forcing to one scalar ($R_1$) can not be strongly correlated with another scalar ($R_2$). Moreover, the small net reaction rate implies that the instantaneous reactive scalar is weakly influenced by other scalar(s). Thus at the statistical stationary state, $\langle\frac{D R_1'}{D t}R_2'\rangle$ can be assumed negligibly small, i.e. $\langle\frac{D R_1'}{D t}R_2'\rangle\sim0$.

As discussed in section~\ref{correlation_derivative}, the correlation coefficients of the reactive scalars is roughly the same as the correlation coefficients of their gradients, which in isotropic turbulence can be estimated as
\begin{align}
  \langle\partial_xR_2'\partial_xR_1'\rangle & \approx \langle\partial_yR_2'\partial_yR_1'\rangle \approx \langle\partial_zR_2'\partial_zR_1'\rangle \nonumber\\
   & \approx \langle(\partial_xR_2')^2\rangle^{1/2} \langle(\partial_xR_1')^2\rangle^{1/2}\cdot\frac{\langle R_2'R_1'\rangle}{\langle R_2'^2\rangle^{1/2} \langle R_1'^2\rangle^{1/2}}\nonumber\\
   & \approx \frac{\langle R_2'^2\rangle^{1/2} \langle R_1'^2\rangle^{1/2}}{\lambda_{\theta}^2}\cdot\frac{\langle R_2'R_1'\rangle}{\langle R_2'^2\rangle^{1/2} \langle R_1'^2\rangle^{1/2}}\approx \frac{\langle R_2'R_1'\rangle}{\lambda_{\theta}^2}, \label{R1_R2_derivative}
\end{align}
where $\lambda_{\theta}$ is the Taylor microscale for scalars.

Consequently,
\begin{equation}\label{R1_R2_derivative_2}
  \frac{1}{Re_{\lambda}Sc}\langle\nabla R_2'\nabla R_1'\rangle=\frac{1}{Re_{\lambda}Sc}(\langle\partial_xR_2'\partial_xR_1'\rangle+\langle\partial_yR_2'\partial_yR_1'\rangle+ \langle\partial_zR_2'\partial_zR_1'\rangle)\approx\frac{3}{Re_{\lambda}Sc\lambda_{\theta}^2}\langle R_2'R_1'\rangle.
\end{equation}
From Eq.~\eqref{eq_R1_R2} it yields
\begin{equation}\label{eq_R1_R2_2}
  3\langle R_2'R_1'\rangle\approx-Da_{\theta} (\langle R_2'R_1'\rangle+n\langle R_2'^2\rangle+\langle R_2'M'\rangle),
\end{equation}
where $Da_{\theta}=Re_{\lambda}Sc\lambda_{\theta}^2Da=\frac{\lambda_{\theta}^2\gamma_1 R_{2,eq}^n}{D}$ is the Damkh\"{o}ler number based on scalar Taylor micro-scale and diffusivity.

Similarly, from Eq.~\eqref{eq_M_R2} and~\eqref{eq_R1_M},
\begin{align}
  3\langle M'R_2'\rangle&\approx-nDa_{\theta} (\langle M'R_1'\rangle+n\langle M'R_2'\rangle+\langle M'^2\rangle),\label{eq_M_R2_2}\\
  3\langle R_1'M'\rangle&\approx-Da_{\theta} (\langle R_1'^2\rangle+n\langle R_1'R_2'\rangle+\langle R_1'M'\rangle).\label{eq_R1_M_2}
\end{align}
Because $R_1'$ and $M'$ are statistically the same (see Eq.~\eqref{eq_scalar_non_dim_1_used_prime_M} and \eqref{eq_scalar_non_dim_3_used_prime_M}), we can define
\begin{equation}\label{C_c_V_v}
  C=\langle R_1'M'\rangle,     \quad       c=\langle R_2'R_1'\rangle=\langle R_2'M'\rangle,      \quad      V=\langle R_1'^2\rangle=\langle M'^2\rangle,     \quad       v=\langle R_2'^2\rangle,
\end{equation}
i.e.
\[\frac{C}{V}=r(R_1,M)=-r(R_1,P),\qquad \frac{c}{\sqrt{Vv}}=r(R_1,R_2)=r(R_2,M)=-r(R_2,P).\]
Then Eq.~\eqref{eq_R1_R2_2},~\eqref{eq_M_R2_2} and~\eqref{eq_R1_M_2} can be rewritten as
\begin{align}
  3c&\approx-Da_{\theta}(2c+nv),\label{C_c_V_v_1}\\
  3c&\approx-nDa_{\theta}(C+nc+V),\label{C_c_V_v_2}\\
  3C&\approx-Da_{\theta}(C+nc+V),\label{C_c_V_v_3}
\end{align}
which then leads to the solutions of Eq.~\eqref{co_R1_P} and~\eqref{co_R1_R2} as
\begin{align*}
  r(R_1,P)=-\frac{C}{V}&\approx\frac{Da_{\theta}}{3+n^2Da_{\theta}+Da_{\theta}},\\
  r(R_1,R_2)=-r(R_2,P)=\frac{c}{\sqrt{Vv}}&\approx\frac{-nDa_{\theta}}{\sqrt{3+n^2Da_{\theta}+Da_{\theta}}\sqrt{3+2Da_{\theta}}}.
\end{align*}

\section{Analytical prediction for reactant variances}\label{theory_fluctuations}
By multiplying Eq.~\eqref{eq_scalar_non_dim_1_used_prime_M}, ~\eqref{eq_scalar_non_dim_2_used_prime_M} and~\eqref{eq_scalar_non_dim_3_used_prime_M} with $R_1'$, $R_2'$ and $M'$, respectively and averaging ($\langle \ldots \rangle$) in space and time, we obtain
\begin{align}
  \frac{1}{2}\frac{D \langle R_1'^2\rangle}{D t}+\frac{1}{Re_{\lambda}Sc}\langle|\nabla R_1'|^2\rangle&\approx-Da (\langle R_1'^2\rangle+n\langle R_1'R_2'\rangle+\langle R_1'M'\rangle)+\langle R_1'\dot{q}_{R_1}\rangle,\label{eq_R1_square_begin}\\
  \frac{1}{2}\frac{D \langle R_2'^2\rangle}{D t}+\frac{1}{Re_{\lambda}Sc}\langle|\nabla R_2'|^2\rangle&\approx-nDa (\langle R_1'R_2'\rangle+n\langle R_2'^2\rangle+\langle R_2'M'\rangle)+\langle R_2'\dot{q}_{R_2}\rangle,\label{eq_R2_square_begin}\\
  \frac{1}{2}\frac{D\langle M'^2\rangle}{D t}+\frac{1}{Re_{\lambda}Sc}\langle|\nabla M'|^2\rangle&\approx-Da (\langle R_1'M'\rangle+n\langle R_2'M'\rangle+\langle M'^2\rangle)+\langle M'\dot{q}_{M}\rangle.\label{eq_M_square_begin}
\end{align}
The dissipation terms above, e.g. $\frac{1}{Re_{\lambda}Sc}\langle|\nabla R_1'|^2\rangle$, can be estimated similarly as Eq.~\eqref{R1_R2_derivative}. Under the isotropic and statistical stationary conditions, the turbulent energy of $R_1'$, $R_2'$ and $P'$ are approximately determined as
\begin{align}
  3\langle R_1'^2\rangle&\approx-Da_{\theta} (\langle R_1'^2\rangle+n\langle R_1'R_2'\rangle+\langle R_1'M'\rangle)+Re_{\lambda}Sc\lambda_{\theta}^2\langle R_1'\dot{q}_{R_1}\rangle,\label{eq_R1_square}\\
  3\langle R_2'^2\rangle&\approx-nDa_{\theta} (\langle R_1'R_2'\rangle+n\langle R_2'^2\rangle+\langle R_2'M'\rangle)+Re_{\lambda}Sc\lambda_{\theta}^2\langle R_2'\dot{q}_{R_2}\rangle,\label{eq_R2_square}\\
  3\langle M'^2\rangle&\approx-Da_{\theta} (\langle R_1'M'\rangle+n\langle R_2'M'\rangle+\langle M'^2\rangle)+Re_{\lambda}Sc\lambda_{\theta}^2\langle M'\dot{q}_M\rangle.\label{eq_M_square}
\end{align}
Similarly, based on Eq.~\eqref{eq_scalar_non_dim_4_used}, the turbulent energy of $T'$ is
\begin{equation}
  3\langle T'^2\rangle\approx Re_{\lambda}Sc\lambda_{\theta}^2\langle T'\dot{q}_{T}\rangle.\label{eq_T_square}
\end{equation}

Define
\begin{equation}\label{W_w_WT_VT}
\begin{split}
  w=Re_{\lambda}Sc\lambda_{\theta}^2\langle R_2'\dot{q}_{R_2}\rangle,  &   \qquad       W=Re_{\lambda}Sc\lambda_{\theta}^2\langle R_1'\dot{q}_{R_1}\rangle=Re_{\lambda}Sc\lambda_{\theta}^2\langle M'\dot{q}_M\rangle,\\
  V_T=\langle T'^2\rangle,   &  \qquad       W_T=Re_{\lambda}Sc\lambda_{\theta}^2\langle T'\dot{q}_{T}\rangle.
\end{split}
\end{equation}

Together with Eq.~\eqref{C_c_V_v},~\eqref{eq_R1_square},~\eqref{eq_R2_square} and~\eqref{eq_T_square} we obtain
\begin{align}
  3V&\approx-Da_{\theta}(V+nc+C)+W,\label{C_c_V_v_W_w_1}\\
  3v&\approx-nDa_{\theta}(2c+v)+w,\label{C_c_V_v_W_w_2}\\
  3V_T&\approx W_T.\label{C_c_V_v_W_w_3}
\end{align}

It is worthy noting that for all the scalar quantities the delta-correlated external forcing is exerted in the same way with constant amplitude. When $\langle R_1'^2\rangle$ is close to $\langle T'^2\rangle$ ($\frac{V}{V_T}$ close to $1$), it is reasonable to assume $W\approx W_T$.
Together with Eq.~\eqref{co_R1_P},~\eqref{co_R1_R2},~\eqref{C_c_V_v_W_w_1} and~\eqref{C_c_V_v_W_w_3} we obtain Eq.~\eqref{R1_P_over_T}:
\begin{equation}
  \frac{\langle R_1'^2\rangle}{\langle T'^2\rangle}=\frac{\langle P'^2\rangle}{\langle T'^2\rangle} = \frac{V}{V_T} \approx \frac{3+n^2Da_{\theta}+Da_{\theta}}{3+n^2Da_{\theta}+2Da_{\theta}}.
\end{equation}
From Eq.~\eqref{C_c_V_v_1}-~\eqref{C_c_V_v_3}, the ratio between the fluctuation magnitudes of $R_2$ and $R_1$ is determined as
\begin{equation}
    \frac{\langle R_2'^2\rangle}{\langle R_1'^2\rangle}=\frac{v}{V}\approx\frac{3+2Da_{\theta}}{3+n^2Da_{\theta}+Da_{\theta}},\label{beta}
\end{equation}
which leads to Eq.~\eqref{R2_over_T}  as
\begin{equation}
  \frac{\langle R_2'^2\rangle}{\langle T'^2\rangle}=\frac{\langle R_1'^2\rangle}{\langle T'^2\rangle}\frac{\langle R_2'^2\rangle}{\langle R_1'^2\rangle}=\frac{V}{V_T}\frac{v}{V} \approx \frac{3+2Da_{\theta}}{3+n^2Da_{\theta}+2Da_{\theta}}.
\end{equation}

\section{On coherency spectrum}\label{correlation_cospectrum}
\textcolor{black}{In the model system at study, characterised by weakly reacting scalars, $R_1$ and $R_2$, sustained by fluctuations  introduced by external sources (here randomly added in the Fourier space).
We find numerically the two following results:
}
\textcolor{black}{
\begin{enumerate}
  \item The energy spectra of $R_1$ and $R_2$ have the same shape;\label{con1}
  \item the correlation coefficient between $R_1$ and $R_2$ conditional on a given wave vector is $k$-independent\label{con2}
\end{enumerate}
These evidences allow to formulate a prediction on the shape of the coherency spectrum as follows.
}
We consider here the one-dimensional case and analysis in three-dimensional space can be implemented similarly. The fluctuating parts of $R_1$ and $R_2$ are expressed in the form of Fourier modes as
\begin{equation}\label{Ak_Bk}
  R_1'=\sum_k A_k(t)\sin (kx)+a_k(t)\cos (kx),\quad R_2'=\sum_k B_k(t)\sin (kx)+b_k(t)\cos(kx).
\end{equation}

Their global correlation coefficient is
\begin{align}
  r(R_1,R_2) = & \frac{\langle [\sum_k A_k(t)\sin (kx)+a_k(t)\cos (kx)][\sum_k B_k(t)\sin (kx)+b_k(t)\cos(kx)]\rangle}  {\langle [\sum_k A_k(t)\sin (kx)+a_k(t)\cos (kx)]^2\rangle^{1/2}\langle[\sum_k B_k(t)\sin (kx)+b_k(t)\cos(kx)]^2\rangle^{1/2}}  \nonumber\\
  = & \frac{ \sum_k \overline{A_kB_k}+\overline{a_kb_k} }  { [\sum_k (\overline{A_k^2}+\overline{a_k^2})]^{1/2}[\sum_k (\overline{B_k^2}+\overline{b_k^2})]^{1/2}},\label{global_correlation}
\end{align}
where $\overline{\cdot}$ denotes time average.

Since the energy distributions of $R_1$ and $R_2$ on different length scales are the same \textcolor{black}{(result 1)}, we define $\alpha$ as
\begin{equation*}
  \frac{\overline{A_1^2}+\overline{a_1^2}}{\overline{B_1^2}+\overline{b_1^2}}=\frac{\overline{A_2^2}+\overline{a_2^2}}{\overline{B_2^2}+\overline{b_2^2}}=\dots=\frac{\sum_k (\overline{A_k^2}+\overline{a_k^2})}{\sum_k (\overline{B_k^2}+\overline{b_k^2})}=\alpha^2.
\end{equation*}
Then the denominator of Eq.~\eqref{global_correlation} can be further written as
\begin{equation*}
  [\sum_k (\overline{A_k^2}+\overline{a_k^2})]^{1/2}[\sum_k (\overline{B_k^2}+\overline{b_k^2})]^{1/2} =  (\sum_k \overline{B_k^2}+\overline{b_k^2})\cdot\alpha = \sum_k \alpha (\overline{B_k^2}+\overline{b_k^2})=\sum_k (\overline{A_k^2}+\overline{a_k^2})^{1/2}(\overline{B_k^2}+\overline{b_k^2})^{1/2}.
\end{equation*}
Thus it yields
\begin{equation}
  r(R_1,R_2) = \frac{ \sum_k \overline{A_kB_k}+\overline{a_kb_k}}  {\sum_k (\overline{A_k^2}+\overline{a_k^2})^{1/2}(\overline{B_k^2}+\overline{b_k^2})^{1/2}}.\label{global_correlation_2}
\end{equation}

The coherency spectrum between $R_1'$ and $R_2'$, $Co_{R_1,R_2}$, describes the correlation coefficients between two scalars corresponding to each length scale. At the mode $k$,
\begin{align}\label{cospectra_scalar_real}
  Co_{R_1,R_2}(k)=&\frac{\langle [ A_k(t)\sin (kx)+a_k(t)\cos (kx)][ B_k(t)\sin (kx)+b_k(t)\cos(kx)]\rangle}  {\langle [A_k(t)\sin (kx)+a_k(t)\cos (kx)]^2\rangle^{1/2}\langle[B_k(t)\sin (kx)+b_k(t)\cos(kx)]^2\rangle^{1/2}}  \nonumber\\
  =&\frac{ \overline{A_kB_k}+\overline{a_kb_k}}  {(\overline{A_k^2}+\overline{a_k^2})^{1/2}(\overline{B_k^2}+\overline{b_k^2})^{1/2}}.
\end{align}
\textcolor{black}{According to the result 2 the correlation coefficients at a given wave-vector between $R_1$ and $R_2$ are independent of the wave-vector, i.e. 
\begin{equation}
\frac{  \overline{A_{k}B_{k}}+\overline{a_{k}b_{k}}}  {(\overline{A_{k}^2}+\overline{a_{k}^2})^{1/2}(\overline{B_{k}^2}+\overline{b_{k}^2})^{1/2}} = const.  \quad \forall k \label{correlation_equal}
\end{equation}
By means of (\ref{cospectra_scalar_real}) and (\ref{correlation_equal}), this gives
\begin{equation}
  Co_{R_1,R_2}(k)=r(R_1,R_2) \quad \forall k.
\end{equation}
}

Therefore, the global correlation coefficient between $R_1$ and $R_2$ is the same as the correlation coefficient at each wave number or length scale.

\begin{figure}[ht]
\centering
\includegraphics[width=0.5\linewidth]{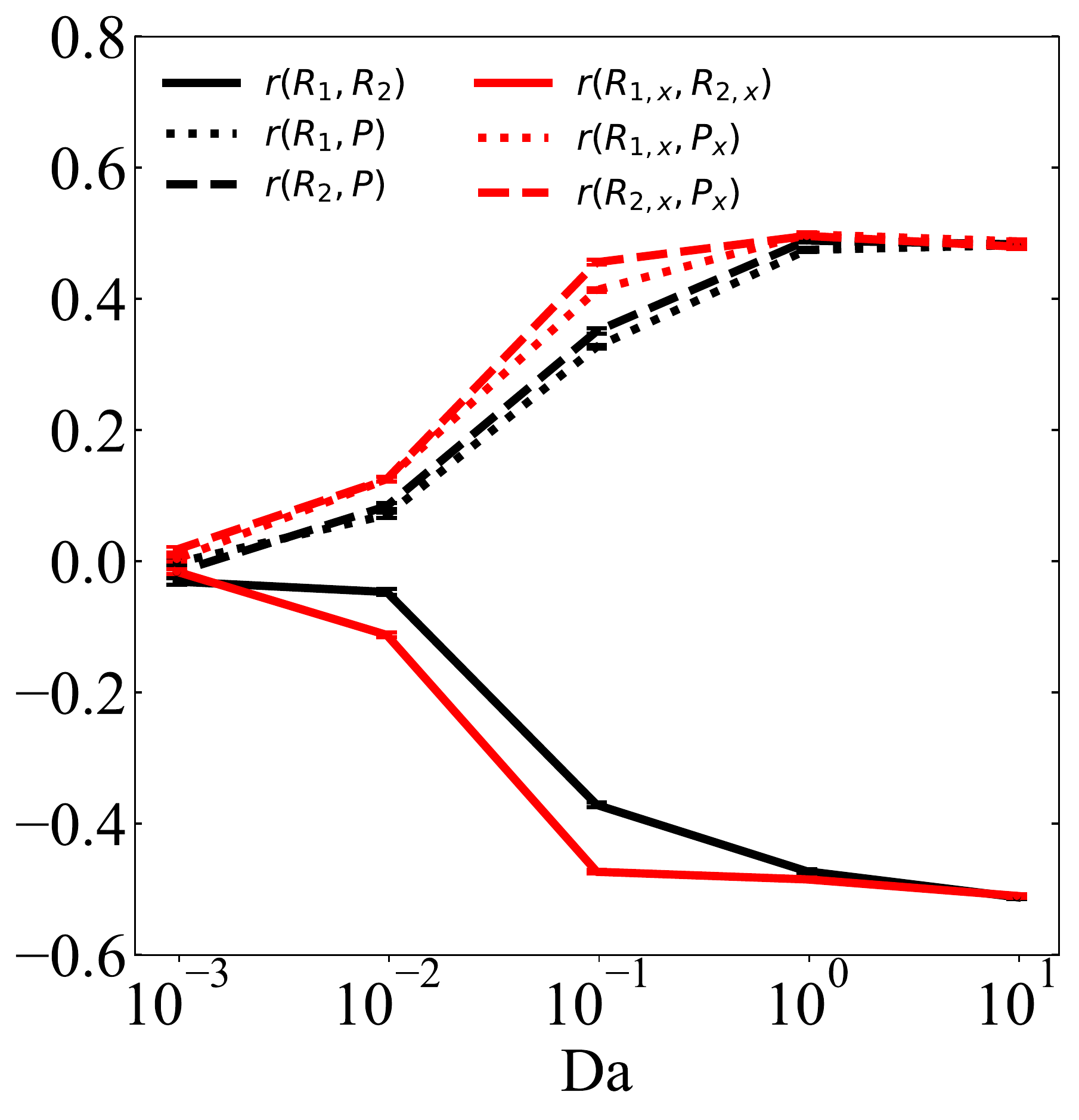}
\caption{The global correlation coefficients of the reactive scalars and their gradients along $x$ direction, under the condition of $Re_{\lambda}=150$, $Sc=1$, $n=1$.}
\label{correlation_ps_and_deri_n_is_1}
\end{figure}

\section{On correlation coefficient of reactive scalar gradients}\label{correlation_derivative}
As shown in Fig.~\ref{ps_spec_n_is_1} and Fig.~\ref{cospec_n_is_1},  the energy distribution and the correlation coefficients of the reactive scalars remain almost independent of the length scale. Under these two conditions, it is ready to derive that the global correlation coefficient between two reactive scalars is the same as their coherency spectrum at each length scale ( as seen in Appendix~\ref{correlation_cospectrum}).

Another quantity of primary importance is the global correlation coefficients of the gradients of scalars, which can be defined along one direction (e.g. $x$) only because of isotropy:
\begin{equation}
  r(\theta_{1,x},\theta_{2,x}) = \frac{\langle \frac{\partial \theta_1}{\partial x}\frac{\partial \theta_2}{\partial x}\rangle}{\langle (\frac{\partial \theta_1}{\partial x})^2\rangle^{1/2}\langle (\frac{\partial \theta_2}{\partial x})^2\rangle^{1/2}}. \label{correlation_x_deri_definition}
\end{equation}
For various scalars, the (almost) identical spectra of the scalar energy implies the (almost) identical spectra of the energy of scalar gradient quantities.
In addition, since Eq.~\eqref{cospectra_scalar} is the definition of the coherency spectrum between not only $\theta_1$ and $\theta_2$ but also their gradients, the coherency spectra between the gradients of two reactive scalars are also almost $k$ independent. Therefore, the correlation coefficient of the gradients of two reactive scalars is also identical at each length scale, and supposed to be the same as the correlation coefficient of these two reactive scalars. Fig.~\ref{correlation_ps_and_deri_n_is_1} presents the global correlation coefficients of the reactive scalars and their gradients against $Da$ with $Re_{\lambda}=150$, $Sc=1$, $n=1$. The speculation that the correlation coefficients between the reactive scalars are the same as that of their gradients is well satisfied, except for the cases of $Da=0.01$ and $0.1$, in which the condition that the coherency spectra is $k$ independent is not satisfied at the largest scales (Fig.~\ref{cospec_n_is_1}).

\begin{acknowledgments}
 This work is under the joint support by Shanghai Jiao Tong University and French Region ``Hauts-de-France'' in the framework of a cotutella PhD programme. We thank Dr. Michael Gauding (CORIA (CNRS UMR 6614), Rouen, France) for providing the code. We acknowledge the computing resources including the High Performance Computing Center (HPCC) at Universit\'{e} de Lille, CALCULCO of Universit\'{e} du Littoral C\^{o}te d'Opale and the National Supercomputer Center in Guangzhou, China.
\end{acknowledgments}


%

\end{document}